\documentclass[preprint]{aastex}
\newcommand{\et}{et\thinspace al.\ }
\newcommand{\kms}{km\thinspace s$^{-1}$}
\newcommand{\vsini}{$v\thinspace sin(i)$}
\newcommand{\arcm}{\ifmmode {' }\else $' $\fi}
\newcommand{\arcs}{\ifmmode {'' }\else $'' $\fi} 
\newcommand{\msini}{$\langle$$sin(i)$$\rangle$}
\newcommand{\mrsini}{$\langle$$R~sin(i)$$\rangle$}
\slugcomment{}

\shortauthors{Rhode et al.}
\shorttitle{Rotational Velocities of ONC Stars}
\begin{document}

\title{Rotational Velocities and Radii of Pre-Main-Sequence Stars
in~the~Orion~Nebula~Cluster}

\author{Katherine L. Rhode\altaffilmark{1}}
\affil{Department of Astronomy, Yale University, New Haven, CT 06520}
\email{rhode@astro.yale.edu}

\author{William Herbst\altaffilmark{1}}
\affil{Astronomy Department, Wesleyan University, Middletown, CT 06459}
\email{bill@astro.wesleyan.edu}

\and

\author{Robert D. Mathieu\altaffilmark{1}}
\affil{Department of Astronomy, University of Wisconsin, Madison, WI 53706}
\email{mathieu@astro.wisc.edu}

\altaffiltext{1}{Visiting Astronomer, Kitt Peak National Observatory,
National Optical Astronomy Observatories, which is operated by the
Association of Universities for Research in Astronomy, Inc. (AURA)
under cooperative agreement with the National Science Foundation}

\begin{abstract}
We have obtained high-dispersion spectra for a sample of 256
pre-main-sequence stars in the Orion Nebula Cluster (ONC) in order to
measure their projected rotational velocities and investigate the
rotational evolution and physical properties of young, low-mass stars.
Half the stars were chosen because they had known photometric periods
and the other half were selected as a control sample of objects
without known periods from the same portion of the H-R diagram.  More
than 90\% of the spectra yielded \vsini\ measurements, although about
one-third are upper limits.  We find strong evidence confirming the
long-held assumption that the periodic light variations of T~Tauri
stars are caused by rotation of spots on their surfaces.  We find no
statistically significant difference between the \vsini\ distributions
of the periodic and control samples, indicating that there is no
strong bias in studying the rotational properties of young stars using
periodic variables.  Likewise, the classical and weak T~Tauri stars
exhibit \vsini\ distributions that are statistically the same.  For
stars with known period and \vsini, the mean value of $sin(i)$ is
significantly lower than expected for a random distribution of stellar
rotation axes.  This could be caused by errors in one or more of the
quantities that contribute to the $sin(i)$ calculation or to a real
physical effect.  We investigate the possible causes and find that
\msini\ has the expected value if we increase the effective
temperatures of our stars by 400$-$600~K. Finally, we have calculated
minimum radii ($R~sin(i)$) for stars with both \vsini\ and period, as
well as average radii for objects grouped by their location in the H-R
diagram.  We find evidence at the three-sigma level that the radii of
the stars on similar mass tracks are decreasing as the stars move
closer to the ZAMS.
\end{abstract}

\keywords{stars: pre-main sequence; stars: rotation; stars:
fundamental parameters}

\section{Introduction}
T~Tauri stars are low-mass, late-type stars that are still undergoing
gravitational contraction toward the main sequence.  They often show
characteristics that reflect the dynamic, turbulent nature of this
early stage of evolution: powerful magnetic fields cause prominent
dark spots to form on their surfaces, bipolar outflows or jets direct
material away from them at high velocities, and accreting matter that
is channelled along magnetic field lines produces surface hot spots
that give rise to rapid brightness fluctuations
\citep{bertout89,herbst94}.  In addition, many T~Tauri stars have thin
circumstellar disks of dust and gas that can extend out to a few
hundred astronomical units.  Studying the rotation and evolution of
these objects can lead to insights into the early evolution of
solar-type stars and the planetary systems that form around them.

One of the best regions of the sky in which to study T~Tauri stars is
the Orion Nebula Cluster (ONC).  The ONC is located in the portion of
the Orion Nebula centered on four OB-type stars called the Trapezium.
The cluster is especially well-suited for investigations of star
formation and early stellar evolution because of its large stellar
population, the range of stellar masses present ($\sim$0.1 -- 50
M$_{\sun}$), and the very young ages ($<$ 1--2 Myr) of its member
stars (Hillenbrand 1997).  In addition, cluster membership for nearly
900 stars in the ONC has been established by a proper motion survey by
\citet{jw88}, and Hillenbrand (1997) and Hillenbrand et al.~(1998)
have published extensive studies of the observable properties of ONC
members.  These reasons, combined with the relative proximity of the
ONC (its distance is $\sim$470 pc; Walker 1969), make it a natural
choice for detailed studies of the physical and rotational properties
of T~Tauri stars.

We began such a study in 1997, with the aim of using high-dispersion
spectroscopy to measure projected rotational velocities, or \vsini\
values, for a large sample of pre-main-sequence stars in the ONC.
Rotation periods, measured via starspot-modulated light curves, have
already been obtained for about 400 stars in the ONC
\citep{herbst00,stassun99,herbst01}.  We sought to combine data
obtained via both photometric and spectroscopic techniques in order to
investigate the fundamental properties and rotational evolution of
young, low-mass stars.  When we began our study, periods for
approximately 130 ONC stars had been measured at Van Vleck Observatory
(VVO) on the campus of Wesleyan University (see Mandel \& Herbst 1991,
Attridge \& Herbst 1992, Choi \& Herbst 1996).  These will hereafter
be referred to as the VVO periods.  Details about the VVO observing
program are described in Herbst et al.~(2000); briefly, multiple
fields around the ONC are imaged repeatedly on each of $\sim$30 clear
nights over the course of an observing season, using a 0.6-meter
telescope and a CCD.  For our \vsini\ study, we chose as targets a
sample of stars with measured VVO periods (hereafter the ``Periodic
sample''), as well as a comparably-sized sample of ONC stars from the
same portion of the H-R diagram {\it without} known rotation periods
(hereafter the ``Control sample'').  The properties of these two
samples are described in more detail in Section~\ref{section:sample
properties}.

In addition to using our \vsini\ measurements to investigate
pre-main-sequence stellar rotation in the ONC, we had other specific
objectives in mind.  First of all, measuring \vsini\ values for stars
with known periods can confirm that rotation of surface cool or hot
spots is indeed the cause of the periodic variations of T~Tauri
stars. It has been assumed since the pioneering studies of Rydgren and
Vrba in the 1980's (e.g., Rydgren \& Vrba 1983, Vrba et al.\ 1986)
that the periodic variations of these objects can be interpreted as
due to rotation.  \citet{stassun99} observed a correlation between
line width and angular velocity among ONC stars, supporting this
assumption.  We address the issue further in this paper.  Also, for
individual stars there is often an ambiguity of a factor of two in the
rotation period (i.e., caused by spots on two sides) as well as the
possibility of aliasing.  Combining \vsini\ and period measures for a
sample of stars of known luminosity and effective temperature can
allow us to test the rotational interpretation of periods for
pre-main-sequence stars and can help us identify cases of period
doubling or aliasing.

Furthermore, for objects with measured rotation periods, we can use
their \vsini\ values to obtain statistical estimates of their radii.
With the exception of four known pre-main-sequence eclipsing binaries
\citep{popper87,casey95,casey98,andersen91,mamajek00,covino00},
%Popper87:EKCep; Casey95,98:TY CrA;Andersen91,Mamajek00:RS Cha
%Covino00:RXJ0529.4+0041
we have no direct measurements of the sizes of pre-main-sequence
stars.  Since $P\thinspace v\thinspace sin(i)$ = $2 \pi R\thinspace
sin(i)$, combining the stellar rotation period $P$ with \vsini\ can
constrain the stellar radius $R$ to within a factor of $sin(i)$.  The
average value of $sin(i)$ can be calculated; thus average radii can be
derived for stars grouped according to their location in the H-R
diagram.  In combination with luminosities, radii also provide an
independent check on effective temperatures, which are typically
derived from spectral types by comparing pre-main-sequence stellar
spectra with those of main sequence or evolved stars.

Another objective of the study was to determine whether the sample of
stars that have persistent spots (and therefore can yield rotation
periods) have rotational properties that are representative of the
properties of ONC members as a whole.  One can imagine that our
understanding of pre-main-sequence rotational evolution might be
biased by data derived from spotted stars.  For example, if
rapidly-rotating stars have more active dynamos, they may be more
likely to have prominent, persistent spots on their surfaces and thus
would be overrepresented in photometric rotation studies.  Comparing
the \vsini\ distributions of the Periodic and Control samples provides
an important observational test of our understanding of the initial
conditions of pre-main-sequence rotation.

Here we present the results from our \vsini\ study of 256 targets in
the ONC.  The following section contains a summary of the
characteristics of the stellar sample and describes the observations
and technique used to measure \vsini. The results of the study are
presented in Section~\ref{section:results}.  In that section, we
explore the measured \vsini\ values of various sub-samples of stars,
compare our \vsini\ values to the equatorial velocities calculated for
stars with measured period, luminosity and effective temperature, and
finally combine period and \vsini\ data to derive average radii for
stars binned according to their location on the H-R diagram.  A
summary of our findings is given in the last section of the paper.

\section{Observations and Analysis}
\label{section:obs and analysis}

\subsection{Characteristics of the Sample}
\label{section:sample properties}
For information about our target stars, we have relied on two previous
studies of the ONC.  Jones \& Walker (1988) undertook a proper motion
study of the cluster and published membership probabilities for 1053
stars in the region.  Throughout this paper, we refer to stars by the
number assigned by Jones \& Walker, e.g., JW~100.  Hillenbrand (1997)
published a comprehensive study of star formation in the ONC,
including optical photometric data for $\sim$1600 stars and spectra
for $\sim$900 stars.  All luminosities, effective temperatures,
spectral types, $I$ magnitudes, and $V$$-$$I$ colors given in this
paper come from the Hillenbrand study.

The initial set of stars observed for this work was selected because
the stars had measured VVO periods.  After the first observing run
(see Section~\ref{section:observations}), a sample of stars without
known periods --- the Control sample --- was chosen.  In order to make
a useful comparison between the \vsini\ distributions of the Periodic
and Control sample stars, the latter were chosen to match the basic
characteristics of the Periodic sample.  The properties of the overall
stellar sample are as follows (the few stars with characteristics
outside those of the larger sample are noted): (1) Spectral types in
the range G6 to M6, with the large majority of stars (88\%) in the
range K0 to M4.  There are three exceptions: JW~945 (spectral type
B6), JW~1033 (spectral type A0$-$A5), and JW~165 (spectral type A7).
(2) $I$ magnitudes from 10$-$16, and $V$$-$$I$ colors ranging from 0.8
to 4.3.  (3) Typical masses (as determined by Hillenbrand (1997) from
pre-main-sequence evolutionary tracks) in the range 0.1 to
3~M$_{\sun}$; the median mass is 0.3~M$_{\sun}$.  (4) Cluster
membership probabilities from Jones \& Walker (1988) equalling 79\% or
above for all but four stars. The large majority of stars (78\%) have
membership probabilities of 99\%. Four stars were observed that have
low membership probabilities; two of them (JW~308 and JW~794) had been
found to be periodic in the VVO study at the time they were observed,
and the other two (JW~3 and JW~1021) have been found to be periodic in
a subsequent study \citep{herbst01}.
%vhelio values: JW3=26.5km/s, JW308=19.8, JW794=124, JW1021=35.1
All four stars exhibit the low-amplitude, sinusoidal light variations
typical of T~Tauri stars.  It is possible that they are T~Tauri stars
within the ONC and that their proper motions are anomalous or in
error, or that they are T~Tauri stars superimposed on the ONC but not
members of the cluster.  For three of them (JW~3, JW~308 and JW~1021)
we measure heliocentric radial velocities that are within 1$-$2
standard deviations from the mean of our distribution (at 27~\kms,
20~\kms\ and 35~\kms, respectively) indicating that these may indeed
be ONC members.  For JW~794 we measure a radial velocity of 124~\kms,
indicating that this may be a true non-member of the cluster.

Because we wished to have roughly equal numbers of objects in the
Periodic and Control samples, and VVO periods were known for $\sim$130
stars, we needed observations of a total of approximately 260
targets. We were able to observe 118 stars for the Periodic sample and
138 for the Control sample.  Table~\ref{table:all data} gives basic
information for the 256 stars observed.  Columns 1 and 3 give the
identification number and membership probability from Jones \& Walker
(1988).  Column 2 specifies whether the star was originally chosen as
part of the Periodic (P) or Control (C) sample. Columns 4, 5, 6 and 7
are, respectively, $I$, $V$$-$$I$, the log of the luminosity, and the
log of the effective temperature for the star from Hillenbrand (1997).
Columns 8 and 9 are the measured near-infrared excess and the
equivalent width of the Ca~II $\lambda$8542 line from Hillenbrand \et\
(1998).  Column 10 is the measured period of variation of the star
from photometric data, and column 11 indicates the reference from
which the period was taken (V = VVO data \citep{herbst00}, S = data
from \citet{stassun99}, E = ESO data from \citet{herbst01}).  Column
12 gives the star's \vsini\ and estimated error measured in this
study, and column 13 lists the projected radius of the star,
calculated by combining the \vsini\ and rotation period (see
Section~\ref{section:radii}).  Column 14 indicates whether the
cross-correlation function showed sufficient structure to indicate
that the star may be a double-lined spectroscopic binary (see
Sections~\ref{section:measure vel} and \ref{section:vsinis} for
details).  The last column in the table specifies when the star was
observed.  Stars that initially showed structure in their
cross-correlation functions were, when possible, observed more than
once.

\begin{figure}
\plotone{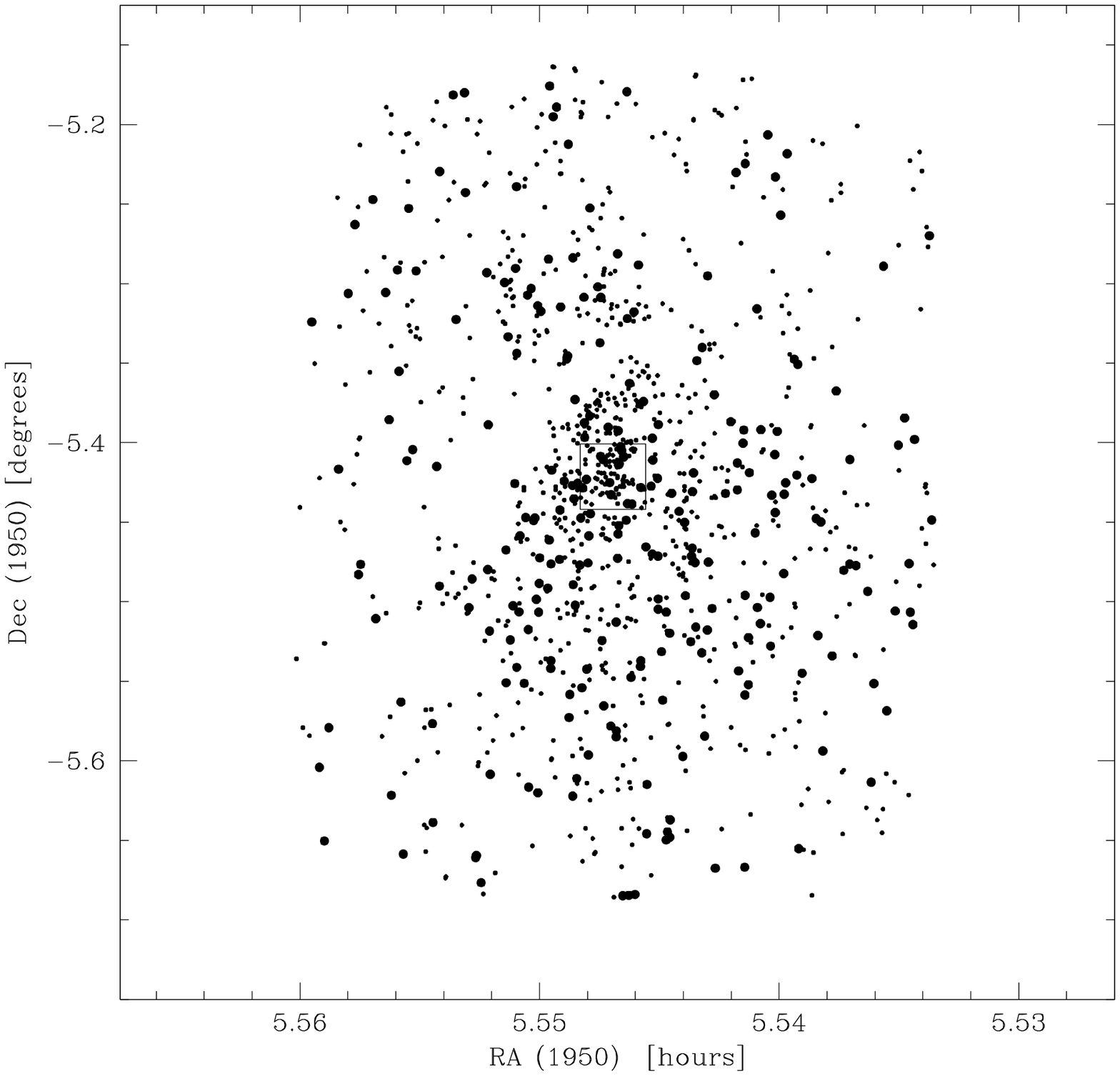}
\caption{The locations on the sky of the 256 ONC stars observed
spectroscopically for this study are shown here as filled circles.
1053 stars included in the Jones \& Walker (1988) proper motion survey
are shown as small dots.  The four Trapezium stars are located at the
center of the small box in the middle of the field.
\protect\label{fig:onc map}}
\end{figure}

The relative spatial locations of the stars observed for this survey
compared to the overall Jones \& Walker (1988) survey are shown in
Figure~\ref{fig:onc map}.  Small dots in the figure show the locations
of the 1053 stars surveyed by Jones \& Walker, and filled circles are
the 256 stars observed for this study.  The observed sample is spread
throughout the area surveyed by Jones \& Walker.  The upper left
portion of the field that is relatively empty of stars is a region of
the ONC that is optically obscured.  Figure~\ref{fig:hrd} shows the
positions of the target stars on an HR diagram.  Stars in the Periodic
sample are shown as filled circles and Control sample stars appear as
open circles.  The pre-main-sequence evolutionary tracks shown in the
figure come from D'Antona \& Mazzitelli (1994, model 1).  The target
stars span the mass range of the tracks and most lie above the
3x10$^6$-year isochrone.  The bulk of the stars are concentrated
toward the low-temperature, low-luminosity end of the diagram.

We note that the Control sample stars were chosen without regard to
whether they appeared within the dozen or so fields that had been
monitored in the VVO program.  This means that they are not
necessarily non-periodic stars (i.e., we were not selecting {\it
against} stars with measurable periods), but are simply ONC members
with the same basic characteristics as the Periodic sample stars.  In
that sense they were selected to be a true ``control'' sample.  Since
this study was initiated, about 30\% of the Control sample stars have
had their periods measured.  Thus Table~\ref{table:all data} contains
stars that are marked with a ``C'' in the Sample column, but also have
a measured period listed.

\subsection{Observations}
\label{section:observations}

Because we needed to obtain high-dispersion spectra for a large number
of objects over a fairly small region of sky (the $_<\atop{^\sim}$1/2
degree subtended by the ONC) a multi-object spectrograph seemed the ideal
instrument for our purposes. The 3.5-meter WIYN telescope\footnote{The
WIYN Observatory is a joint facility of the University of
Wisconsin-Madison, Indiana University, Yale University, and the
National Optical Astronomy Observatories.} at Kitt Peak National
Observatory is equipped with the Hydra Multi-Object Spectrograph, which
makes possible simultaneous spectroscopy of $\sim$100 targets over a
1-degree field.  Observations for this study took place at WIYN during
three separate observing runs, in January 1997, December 1997 and
December 1998. 

For all observations we used the Bench Camera, Red Fiber cable, and
$316@63.4$ Echelle grating.  This spectrograph setup gave us fairly
high resolution ($R$ $\sim$ 21,500) while still allowing us to observe
our faintest targets ($V$ $\sim$ 18) with reasonable integration
times.  The spectra cover the region 6240 \AA\ to 6540 \AA, with
$\lambda_{central}$ = 6400 \AA, and a dispersion of 0.145 \AA\ per
pixel. We chose to work in the red portion of the spectrum because of
the colors of our program stars ($V$$-$$I$ = 1$-$4), while at the same
time avoiding the region of the spectrum around H$\alpha$, where
nebular emission can easily overwhelm the light from the star.  The
spectral resolution (corresponding to a typical FWHM of the slit
profile of 2 pixels) is 13.6~\kms\ at 6400~\AA.

The 256 target stars were observed in five different fiber
configurations, so typical configurations had fibers positioned on
$\sim$50$-$60 objects.  Crowding in the ONC and limitations on exactly
how Hydra fibers can be positioned prevented us from observing larger
numbers of stars in a given configuration.  At least 15$-$20 of the
remaining fibers in each pointing were placed on blank sky, to be used
for sky subtraction during processing of the spectra.  Total
integration times for the program star fields ranged from $\sim$3$-$4
hours, depending on the magnitudes of the stars in the configuration
and the sky conditions.  In addition to the program stars, we observed
five objects from the Gliese Catalog of Nearby Stars \citep{gliese91}
with spectral types spanning the range of our targets; the spectra of
these stars were used as narrow-lined templates in the
cross-correlation process.  Information about the templates is given
in Table~\ref{table:templates}. Daytime sky spectra were also obtained
for use as G2V template spectra.

\begin{figure}
\plotone{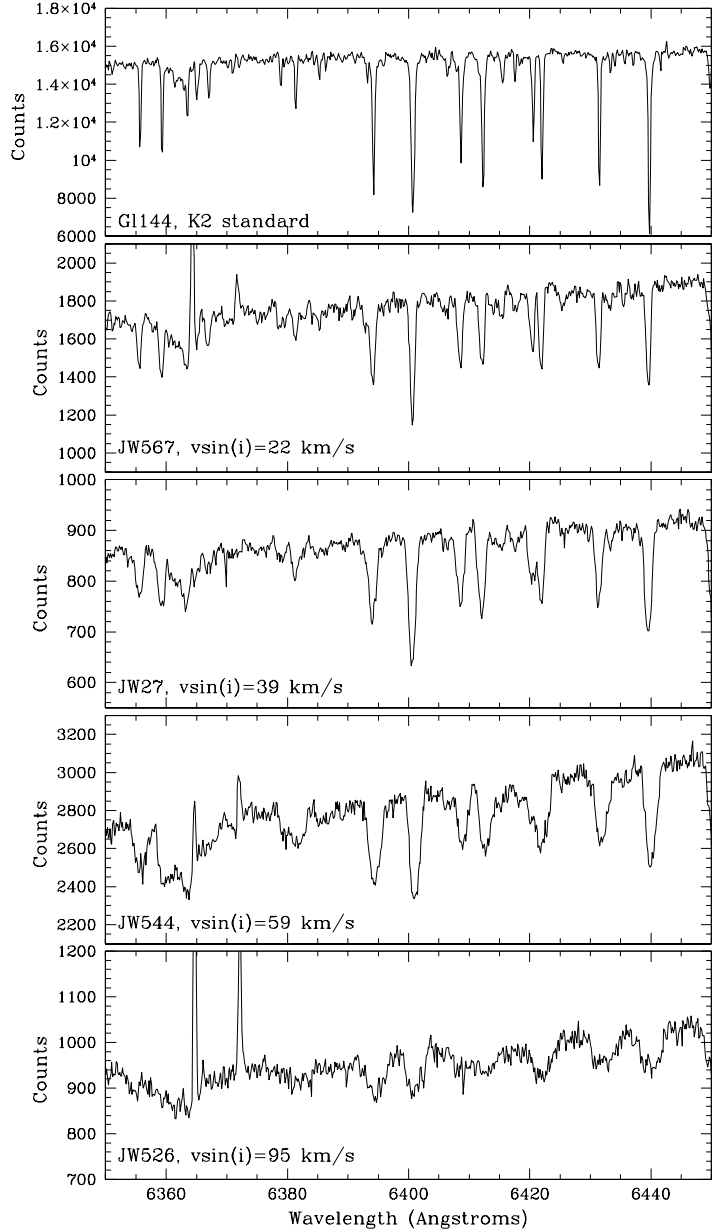}
\caption{Examples of template star and target star spectra obtained
for this study.  The top spectrum is from Gliese 144, a K2V star used
as a spectral template in the cross-correlation procedure.  Its
rotational velocity is below our \vsini\ measurement limit.  The
bottom four panels show spectra of Jones \& Walker (1988) stars with
varying projected rotational velocities; the stellar absorption lines
become broader and more shallow with increasing \vsini. Emission lines
from the Orion Nebula appear in some of the spectra.
\protect\label{fig:spectra}}
\end{figure}

Figure~\ref{fig:spectra} shows examples of narrow-lined template and
object spectra obtained for this study.  The spectra are arranged in
order of increasing rotational velocity, and illustrate the effect
that this has on stellar absorption lines. The top spectrum is that of
Gliese 144, a K2 dwarf used as a narrow-lined template.  Its
rotational velocity is below our \vsini\ measurement limit, and its
lines appear narrow.  The spectra in the bottom four panels are target
stars with projected velocities ranging from $\sim$20$-$100 \kms.  The
spectral lines become broader and shallower with larger \vsini.  The
emission lines present in a few of the spectra are nebular lines at
$\lambda$6363.8 due to [O~{\small I}] and $\lambda$6371.3 due to
Si~{\small II}.  Regions of the spectra with emission lines were
excluded from the \vsini\ measurement process (i.e., not included in
the cross-correlation) described in Section~\ref{section:measure vel}.

\subsection{Initial Reductions}
\label{section:reductions}
%Steps used were: setinst, zerocombine, ccdproc, flatcombine, then dohydra

Preliminary data reduction was accomplished with standard
IRAF\footnote{IRAF is distributed by the National Optical Astronomy
Observatories, which are operated by the Association of Universities
for Research in Astronomy, Inc., under cooperative agreement with the
National Science Foundation.} tasks, e.g., ZEROCOMBINE to create
combined bias images, CCDPROC to bias-subtract and trim the object
images, dome flats and comparison lamp frames, and FLATCOMBINE to
create combined dome flats.  Then the IRAF task DOHYDRA was used to
perform spectral extraction, flat-fielding, fiber throughput
correction, wavelength calibration, and sky subtraction on the object
frames.  The wavelength calibration was done using observations of a
Thorium-Argon comparison lamp.  In the sky subtraction step, sky
fibers with unusually high signal (due to contamination by an object)
were identified and deleted.  A combined sky spectrum was then made by
averaging the remaining sky spectra and using a sigma-clipping
algorithm to eliminate cosmic rays.  We note that the sky fibers
placed randomly around the field will collect flux from both the sky
and the Orion nebula.  The sky values varied spatially, reflecting
variations in the nebular continuum.  Although the sky level in the
vast majority of fibers varied by $_<\atop{^\sim}$10\%, a few had
deviations of as much as 30\%.  A series of tests was performed that
indicate that a variable background does not affect the final measured
\vsini\ values.  The template star and daytime sky spectra were
reduced in the same manner as the target star spectra, with the only
difference being that sky subtraction was not executed on the daytime
sky frames.

Once DOHYDRA had been run on the object frames, individual
integrations taken of the same field (within a given observing run)
were scaled and then combined into a single frame using the task
SCOMBINE, to remove cosmic rays and produce a single, high
signal-to-noise observation.  Because the signal degrades at the ends
of some of the images, the frames were clipped so that the spectra
cover the region between 6275 and 6525 \AA.  All stellar spectra were
continuum-normalized before the cross-correlation step was executed.

\subsection{Measuring Radial and Rotational Velocities}
\label{section:measure vel}

Rotational and radial velocities were derived using the IRAF task
FXCOR, which performs Fourier cross-correlation of an object spectrum
against a specified template.  Figure~\ref{fig:spectra} shows that
increasing rotational velocity causes stellar absorption lines to
become broader and more shallow.  When a stellar spectrum with
rotationally-broadened lines is cross-correlated against a
narrow-lined spectrum, the width of the cross-correlation peak is
sensitive to the amount of broadening present in the first spectrum.
Thus, by measuring the width of the peak, one can obtain a measurement
of the rotational velocity of the star.

\begin{figure}
\plotone{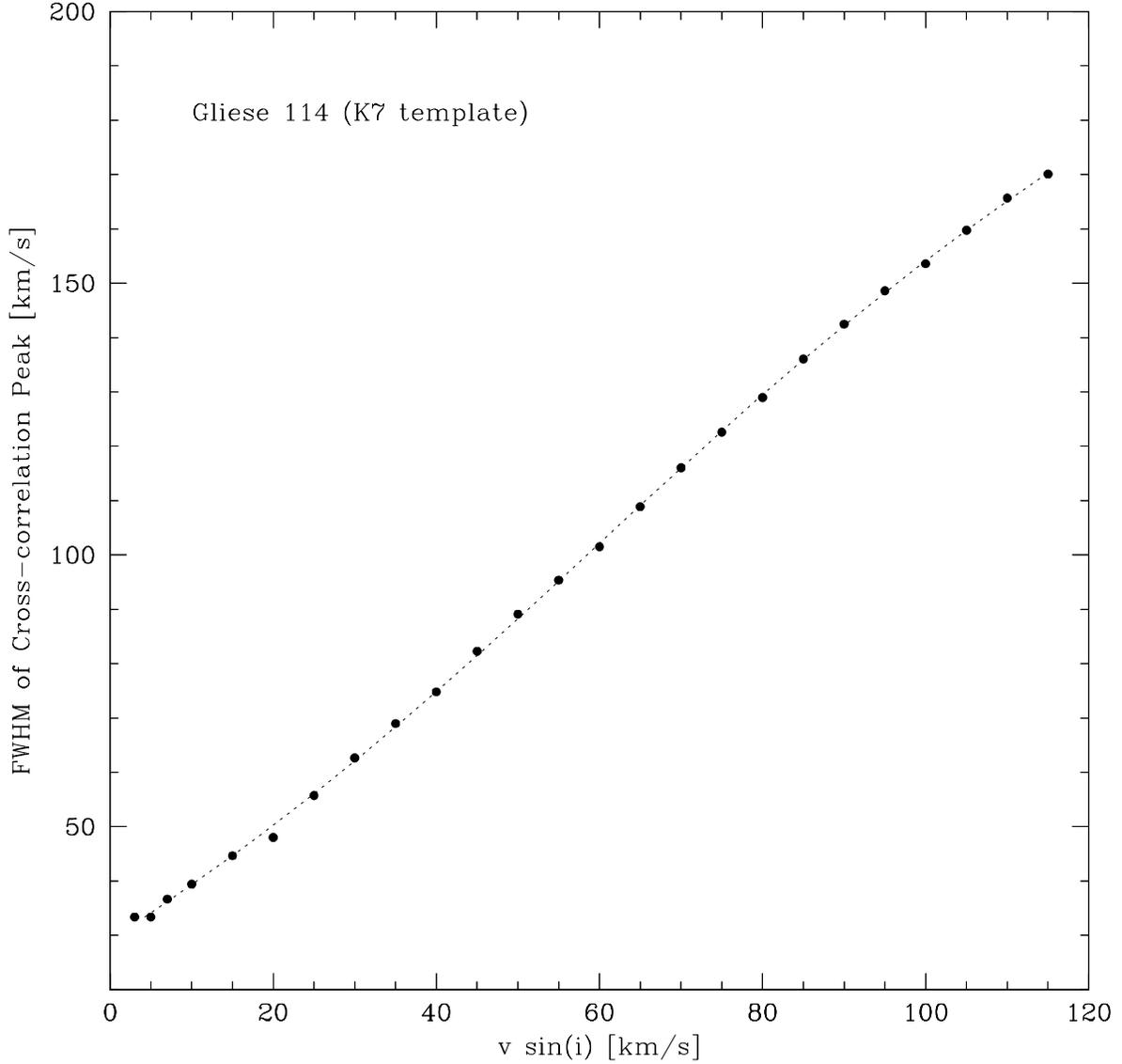}
\caption{Calibration curve for the narrow-lined template star
Gliese~114, created by cross-correlating the original spectrum of this
star against a series of artificially-broadened versions of the
spectrum.  The FWHM of the cross-correlation peak was measured and is
plotted here against the velocity corresponding to each broadened
spectrum (points). The polynomial fit to the FWHM vs. velocity data is
shown as a dotted line. The relationship between the width of the peak
and \vsini\ is fairly linear above 7~\kms.
\protect\label{fig:calib curve}}
\end{figure}

To measure \vsini\ for our target stars, we began by calibrating the
relationship between the width of the cross-correlation peak and
\vsini.  To do this, a narrow-lined template spectrum was artificially
``spun up'' to mimic that of a higher-velocity star by convolving it
with a theoretical rotation profile \citep{gray92}.  A series of such
broadened spectra was created for \vsini\ values spanning the expected
range of observed values.  Each broadened spectrum was then
cross-correlated against the original narrow-lined spectrum, and the
FWHM of the cross-correlation peak was measured.  The resultant
relationship between \vsini\ and the FWHM of the cross-correlation
peak was fit with a 4th-order polynomial.  This process was executed
for each of the six template stars (including the Sun).
Figure~\ref{fig:calib curve} shows, as an example, the calibration
curve data produced by cross-correlating broadened versions of the
spectrum of the template star Gliese~114 against the unbroadened
spectrum.  The points on the figure are the data and the dotted line
is the polynomial fit to the data.  Below 7~\kms, the relationship
between the FWHM of the cross-correlation peak and \vsini\ is flat,
but from 7~\kms\ upward it is fairly linear.  Figure~\ref{fig:calib
curve} illustrates the absolute lower limit on a \vsini\ measurement
using our analysis method, under ideal circumstances.

The actual limit to which we can measure \vsini\ depends in part on
the size of the slit (fiber) image.  With the Bench Camera and our
spectrograph setup, the slit image width (as determined by measuring
the FWHM of emission line profiles in the comparison lamp spectra)
varied from, typically, 1.5 to 1.7 pixels toward the ends of the
spectra to $\geq$2 pixels in the middle (in the worst cases, small
portions of some of the spectra have FWHM as large as 2.8$-$2.9
pixels). This makes determining the exact limit on \vsini\ measurement
more difficult than if the focus were uniform across the CCD chip.  In
the regime in which the broadening of a spectral line can be measured
(i.e., once the width due to intrinsic broadening is larger than the
slit width), \vsini\ is related to the half-width at zero intensity
(HWZI) of the line.  The base of the line, or full-width at zero
intensity (FWZI), is on the order of two times the FWHM of the line,
which means that HWZI is approximately equal to FWHM.  The FWHM of the
slit image varies in our spectra from $<$2 pixels to $>$2 pixels.
Spectral lines from all of these regions contribute to the
cross-correlation.  In places where, for example, the FWHM is 1.7
pixels, the base of a spectral line spans 3$-$4 pixels and therefore
FWZI is resolved.  In such circumstances \vsini\ values somewhat below
the 2-pixel-FWHM spectral resolution of 13.6~\kms\ can be measured.

Because of the focus variation we cannot determine with certainty what
the limit on measuring \vsini\ is on a star-by-star basis. Our
calibration curve suggests that we are sensitive to rotational
broadening as small as 7~\kms.  However, without observations of
\vsini\ standards in that regime, we are not comfortable reporting
\vsini\ measurements much below our spectral resolution.  Thus we take
our limit on \vsini\ measurement to be 11~\kms, since this corresponds
to a best-focus value of between 1.6 and 1.7 pixels (FWHM).  We
emphasize that reported values at or somewhat above this limit should
be treated with caution.  In some cases these values could be upper
limits rather than actual velocity measurements.

Once the relationship between the width of the cross-correlation peak
and \vsini\ was established for the narrow-lined templates, each
program star spectrum was cross-correlated against one of the template
spectra.  The template star closest in $T_{\rm eff}$ to the object was
used as the cross-correlation template.  The FWHM of the
cross-correlation peak was measured and the \vsini\ value for the
target star calculated using the derived relationship between those
two quantities.  Radial velocities for the target stars were obtained
as part of the cross-correlation step.  Radial velocities used for the
template stars come from the Gliese catalog \citep{gliese91}.

\begin{figure}
\plotone{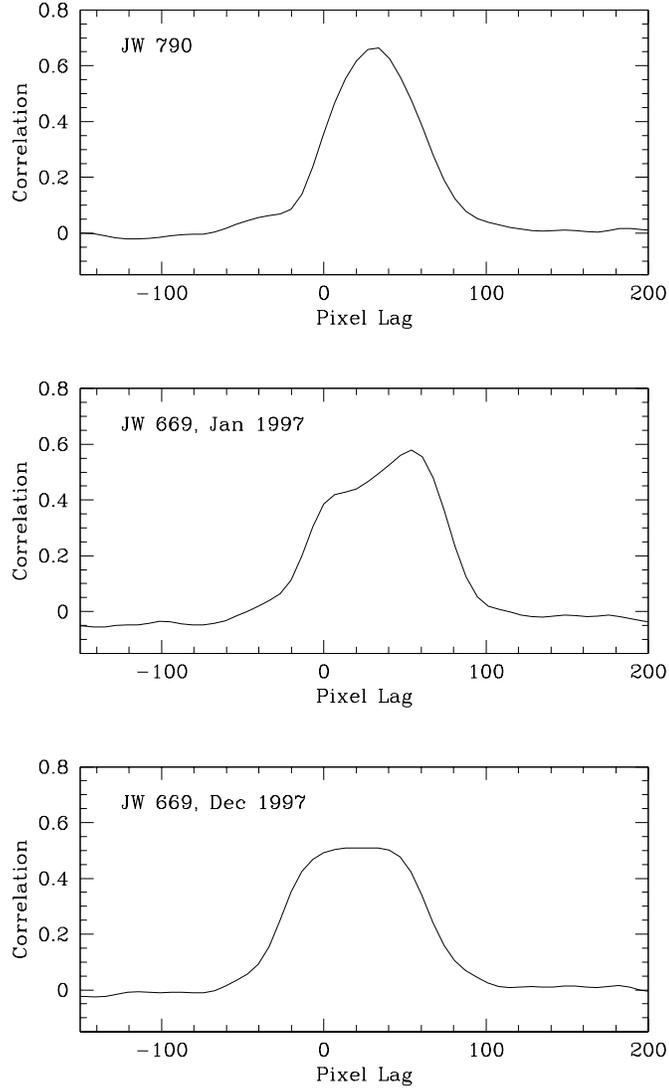}
\caption{Shown here are the peaks of the cross-correlation functions
for the target stars JW~790 (top panel) and JW~669 (bottom two
panels).  The cross-correlation functions are produced by
cross-correlating the stars against the narrow-lined template star
closest in effective temperature to the target.  For JW~790, the
cross-correlation peak is regular in shape and well-approximated by a
Gaussian function.  The peaks for JW~669, made using spectra taken in
January 1997 and December 1997, appear more like two overlapping
Gaussians, an indication that this star may be a spectroscopic
binary.
\protect\label{fig:three ccfs}}
\end{figure}

In addition to being useful for measuring radial and rotational
velocities, the cross-correlation function can also show structure
that may indicate that a star is a double-lined spectroscopic binary
(SB2). The peak of the function, instead of appearing Gaussian in
shape, may look like two Gaussians superimposed on each other, due to
the presence of two sets of lines in the stellar spectrum.
Figure~\ref{fig:three ccfs} shows example cross-correlation functions
for two stars.  The top panel of the figure is the peak of the
cross-correlation function for JW~790.  The peak is fairly regular and
nearly Gaussian in shape.  The bottom two panels are the
cross-correlation function peaks for JW~669, made using spectra from
January 1997 and December 1997.  The structure in this star's
cross-correlation peak indicates that it may be an SB2.

\section{Results}
\label{section:results}

\subsection{Rotational Velocities}
\label{section:vsinis}

Of the 256 stars
%(118 Periodic and 138 Control)
observed for this study, 18 
%(7 from the Periodic sample and 11 from the Control sample) 
did not yield reliable \vsini\ values, primarily because their spectra
lacked sufficient signal.  In these cases, there was either no obvious
peak present in the cross-correlation function, or the peak found by
the FXCOR task was not far enough above the noise to be believable.
Eighty-three of the remaining 238 stars have \vsini\ less than or
equal to our estimated limit of 11~\kms.  The measured \vsini\ values
for our program stars are given in Table~\ref{table:all data}.  For
the 18 stars for which no \vsini\ was measured, ``........'' is given
in the \vsini\ column; for stars with \vsini\ below our estimated
limit, ``$\leq$11.0'' is given in that column.

Seven stars
%(JW~50, JW~99, JW~122, JW~440, JW~669, JW~883, and JW~961)
had sufficient structure in their cross-correlation functions to
indicate that they may be SB2s.  The cross-correlation peaks for these
stars are shown in Figures~\ref{fig:three ccfs} and \ref{fig:six
ccfs}.  Choosing SB2 candidates was a fairly subjective process; we
simply examined the cross-correlation peaks and noted those that were
shaped more like two (or more) Gaussians superimposed than like a
single, smooth Gaussian curve.  Other stars also showed structure in
their cross-correlation peaks, in the sense that the peaks appeared
noisy and irregular rather than smooth, but we did not mark a star as
an SB2 candidate unless its peak looked like multiple Gaussians.  The
seven SB2 candidates are indicated in Table~\ref{table:all data} with
a ``Y'' in the ``SB2?''  column.  If any of them are in fact SB2s,
their \vsini\ values would likely be too large, since some of the width
of the cross-correlation peak would be due to the fact that their
spectra contain two sets of lines.

We have calculated uncertainties on our \vsini\ values by making use
of the parameter $r$ from \citet{tonry79}. It provides a measure of
the signal-to-noise of the cross-correlation peak; specifically, $r$
is the ratio of the height of the cross-correlation peak to that of
the average peak in the noise component of the cross-correlation
function.  \citet{tonry79} showed that the errors in velocities
determined via cross-correlation should be proportional to
($1~+~r$)$^{-1}$.  \citet{hartmann86} used cross-correlation to
measure \vsini\ for stars in Taurus-Auriga and Orion, and investigated
their observed errors as a function of $r$.  They found that the
quantity $\pm$\vsini/$(1 + r)$ provided a good estimate for the 90\%
confidence level of one of their \vsini\ measurements.

We have adopted the value $\pm$\vsini/$(1 + r)$ as a {\it reasonable
estimate} of the one-sigma errors on our \vsini\ values.  We can check
this by comparing \vsini\ values for the eight stars for which we have
repeated observations.  Two of the stars, JW~2 and JW~433, have
\vsini\ values below our estimated measurement limit.  For the
remaining six stars, the percentage difference between the \vsini\
measurements from different observing runs ranges from 4\% to 24\%,
with an average difference of 13\%.
%median difference is 14%
(We note that two of these six stars, JW~99 and JW~669, are SB2
candidates, so the widths of their cross-correlation peaks may change
due to orbital motion of the member stars, if they are indeed
binaries.  However, the percentage differences between the \vsini\
values from their observations, at 14\% and 4\%, respectively, are
consistent with the differences for the other stars.)  The relative
\vsini\ errors calculated using the $r$ parameter for the six stars
range from 10\% to 26\%, with an average error of 20\%.
%median is 21%
Furthermore, when we compare the observed percentage difference in
\vsini\ to the relative \vsini\ errors calculated from $r$ on a
star-by-star basis, we find that the uncertainty calculated from $r$
is equal to or larger than the observed difference for all six of the
stars.  Thus the average error calculated from $r$ is comparable to or
larger than our estimated precision based on repeated measurements.
The quantity $\pm$\vsini/$(1 + r)$ is given as the error on \vsini\ in
Table~\ref{table:all data}.  For the stars with repeated measurements,
the \vsini\ and error given in Table~\ref{table:all data} are the
weighted mean and its associated error, calculated from the multiple
measurements.

In the absence of a larger number of cases in which stars had repeated
observations taken during separate observing runs, we can perform
another test of whether our \vsini\ error estimates are appropriate.
As explained in Section~\ref{section:reductions}, multiple
observations of a given Hydra field taken during the same observing
run were combined to remove cosmic rays and create a single, high
signal-to-noise observation.  A useful comparison to make is to take
these individual Hydra frames and measure \vsini\ for the stars
appearing in them.  This is equivalent to having multiple, independent
observations of the stars in these fields.  Moreover, the
signal-to-noise ratios of the spectra in each frame will differ
depending on the observing conditions during the individual
integrations, providing an additional test of how well our \vsini\
measurements and errors hold up under varying levels of
signal-to-noise.  To perform this test, we analyzed Hydra frames taken
of two different fields, one observed during December 1997 and another
in December 1998.  In the former case, six individual integrations of
the same field were taken over a period of about three weeks.  In the
latter case, five exposures of another field were obtained on a single
night.  To measure \vsini\ in the single frames, we used the identical
method described in Section~\ref{section:measure vel} used to measure
velocities in the combined frames.  Because the integration times for
the single frames were only 1/5 to 1/6 of the total combined
integration, many of the stars in the single frames had insufficient
signal to measure \vsini.  Sixty-one objects did yield reliable
multiple measurements, and of these, 34 have \vsini\ $>$ 11~\kms.  For
these 34 stars, we compared their calculated \vsini\ errors
($\pm$\vsini/$(1 + r)$) with the standard deviation of the individual
\vsini\ values from the multiple measurements.  For 31 of 34 stars
(91\%), the \vsini\ error was equal to or larger than the standard
deviation of the individual measurements, and in a handful of cases
much larger.  For the remaining three stars, the computed \vsini\
error equalled $\sim$60$-$80\% of the standard deviation.  Overall the
test indicates again that our adoption of $\pm$\vsini/$(1 + r)$ as a
reasonable estimate of the one-sigma error on the \vsini\ values is
both valid and appropriate.

A possible source of systematic error in the \vsini\ measurement
process comes from template mismatch, i.e., an error that may arise if
an object spectrum is cross-correlated against a template star that
does not exactly match the object's spectral type.  Template mismatch
might be a factor in our data set because we do not have a template
star to match each individual spectral type of our objects, but
instead have a set of five templates spanning the approximate range of
our targets.  To investigate the effect of template mismatch, a set of
40 stars was randomly chosen from the 238 for which \vsini\ was
successfully measured.  These 40 stars were cross-correlated against
template stars hotter than and cooler than the correct template based
on the star's spectral type.  For example, a K6 target that was
originally cross-correlated against the K7V template star Gliese~114
was for this test cross-correlated against Gliese~144 (K2V) and
Gliese~15A (M1.5V).  Comparisons were made of the \vsini\
values resulting from these tests for the 27 stars with \vsini\ values
of at least 11~\kms.  In 22 of the 27 cases (82\%), the \vsini\ values
measured using the hotter or cooler templates matched the original
\vsini\ within the one-sigma error bars, and in all 27 cases they
matched within less than three sigma.  The mean difference between the
original \vsini\ value and that measured using a hotter template was
$-$2.5~\kms, with a standard deviation of 3.9~\kms.  The mean
difference between the original value and the value when a cooler
template was used was $+$1.3~\kms, with standard deviation 5.1~\kms.
In other words, using a template that is too hot for a given target
star causes the \vsini\ measured to be too large, and the opposite is
true for a template that is too cool.  In the case of the too-cool
template, the mean difference is not statistically significant,
whereas it is marginally so for the templates that were too hot.  We
note however that the test performed mimics an extreme situation in
which the stars have been cross-correlated against templates that are
several spectral subclasses away from the appropriate spectral type.
Even under such circumstances, the mean systematic difference for the
group of 27 stars is very small.  The overall agreement within the
errors for individual stars indicates that the uncertainty associated
with template mismatch is accounted for by our estimated errors on
\vsini.

Another potential source of error in our \vsini\ values is line
blending in the stellar spectra.  Including regions of the spectra
that have strong line blending could bias our results toward larger
derived \vsini.  To test for this effect, 40 stars were randomly
chosen (a different set than used for the template mismatch test) from
the 238 with measured \vsini.  Several regions with possible line
blending were identified using a high signal-to-noise daytime sky
spectrum.  These regions were then excluded from the cross-correlation
process.  The resultant \vsini\ values for 28 stars for which \vsini\
was at least 11~\kms\ were compared with those originally measured.
For 27 of the 28 stars (96\%), the \vsini\ values from the test were
the same as the original values within two sigma, and in all 28 cases
they agreed within less than three sigma.  The mean difference between
the original \vsini\ and that measured with line-blended regions
excluded was 0.45~\kms, with a standard deviation of 3.0~\kms.  This
difference is not significant.  Thus, as in the case of template
mismatch, we again find that the potential systematic uncertainties
associated with line blending are taken into account by the
uncertainties we have estimated for our \vsini\ values.

\begin{figure}
\plotone{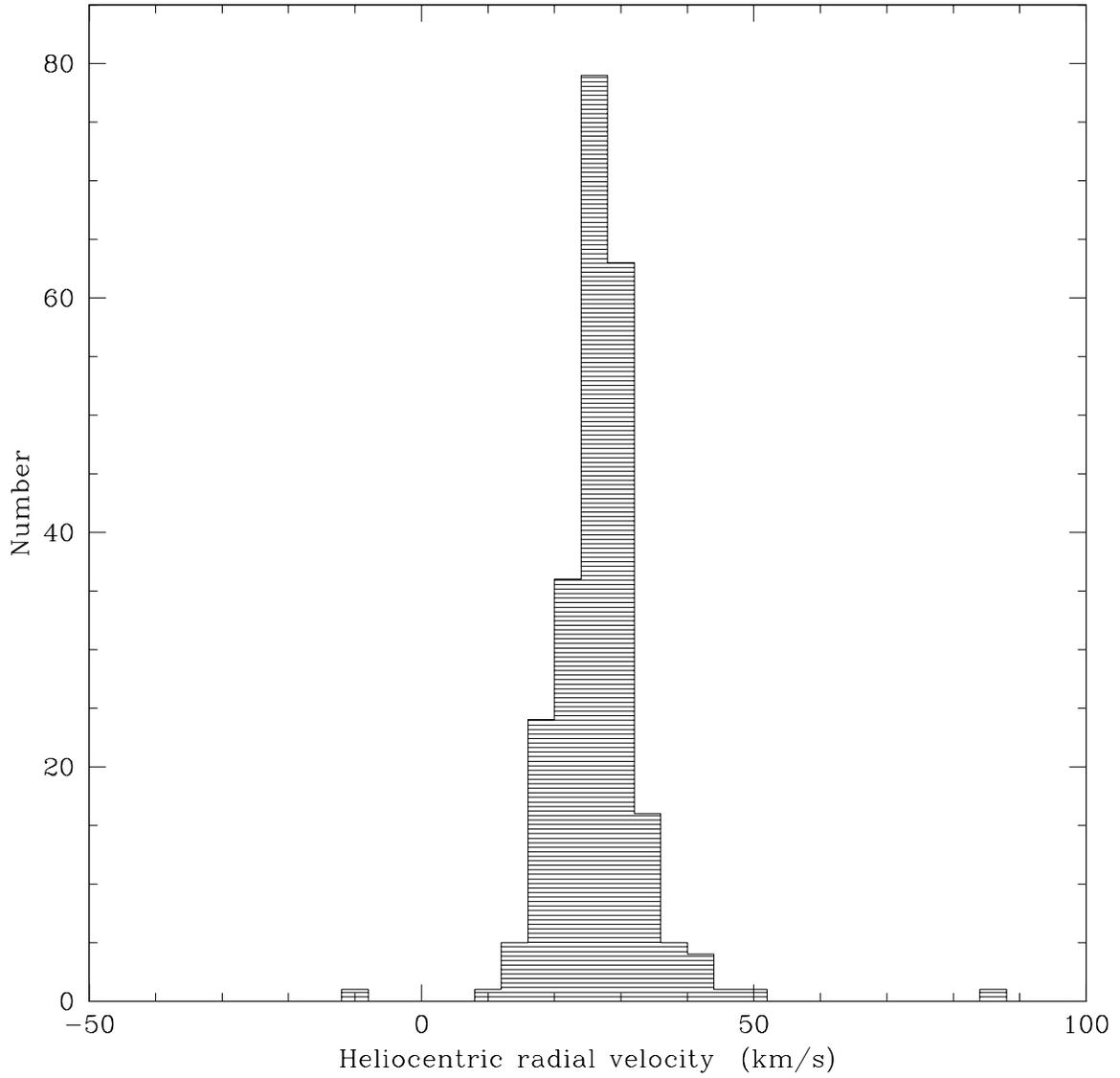}
\caption{Histogram of the heliocentric radial velocity values for the
stars for which \vsini\ was measured.  The mean and standard deviation
of the distribution are 26.7~\kms\ and 5.6~\kms, respectively.
\protect\label{fig:vhelio}}
\end{figure}

Figure~\ref{fig:vhelio} shows a histogram of the heliocentric radial
velocity values for the stars for which \vsini\ was measured.  The
outlier appearing at $-$8 \kms\ is JW~50, one of our SB2 candidates;
to derive the radial velocity for this star, we (arbitrarily) chose to
fit the blue peak in the cross-correlation function, since the two
peaks were well-separated.  The star with a radial velocity of 87
\kms\ is JW~363, which has an ONC membership probability from
\citet{jw88} of 98\%, so this star may be an SB1.  The mean of the
distribution, excluding JW~363 and all SB2 candidates, is 26.7~\kms,
with a standard deviation of 5.6~\kms.  This is comparable (although
with larger errors) to the expected mean radial velocity for stars in
this region, which is $\sim$25~\kms\ with a dispersion of
$\sim$2~\kms\ (Jones \& Walker 1988; R. Mathieu, unpublished data).
%Stassun et al.'s radial velocity mean was 25.7 with rms 3.7

\begin{figure}
\plotone{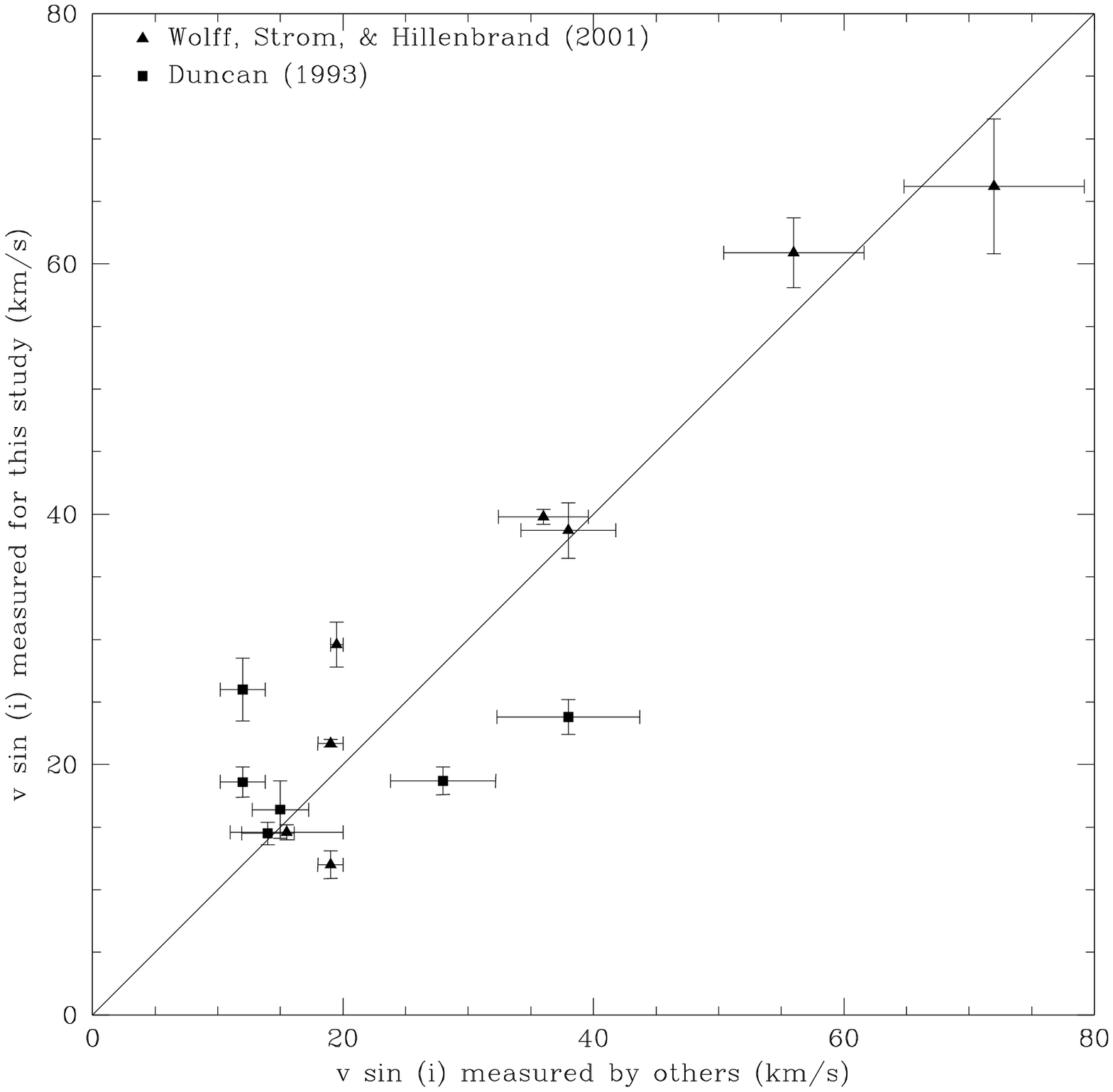}
\caption{Comparison of \vsini\ values for 14 stars observed for this
study and values from \citet{duncan93} (squares) and
\citet{wsh01} (triangles).
\protect\label{fig:compare vsini}}
\end{figure}

Fourteen of the stars in our sample with \vsini\ $>$11~\kms\ have had
their \vsini\ values measured in other studies.
Figure~\ref{fig:compare vsini} shows a comparison between our measured
\vsini\ values and those from work by \citet{duncan93} and
\citet{wsh01}.  In the figure, squares represent stars that our study
has in common with Duncan's study, and triangles are stars we have in
common with Wolff, Strom, \& Hillenbrand.  Duncan estimated that his
\vsini\ values have an accuracy of 10$-$15\%, so the squares are shown
with horizontal error bars equal to 15\% of \vsini.  The points in the
figure scatter around the line marking equal \vsini\ values from our
study and the others; our measured values are larger for nine of the
stars, and smaller for the remaining five.  The mean difference
between the \vsini\ values is 0.5$\pm$2.0~\kms, indicating reasonable
agreement between our values and those from previous work.  The figure
does show, however, that the random errors of all three studies have
probably been underestimated, since about half of the points do not
intersect the equality line within the error bars.  Nevertheless,
there are no obvious systematic differences between our \vsini\ values
and those measured by other groups.

\subsubsection{\vsini\ Distributions}

\begin{figure}
\plotone{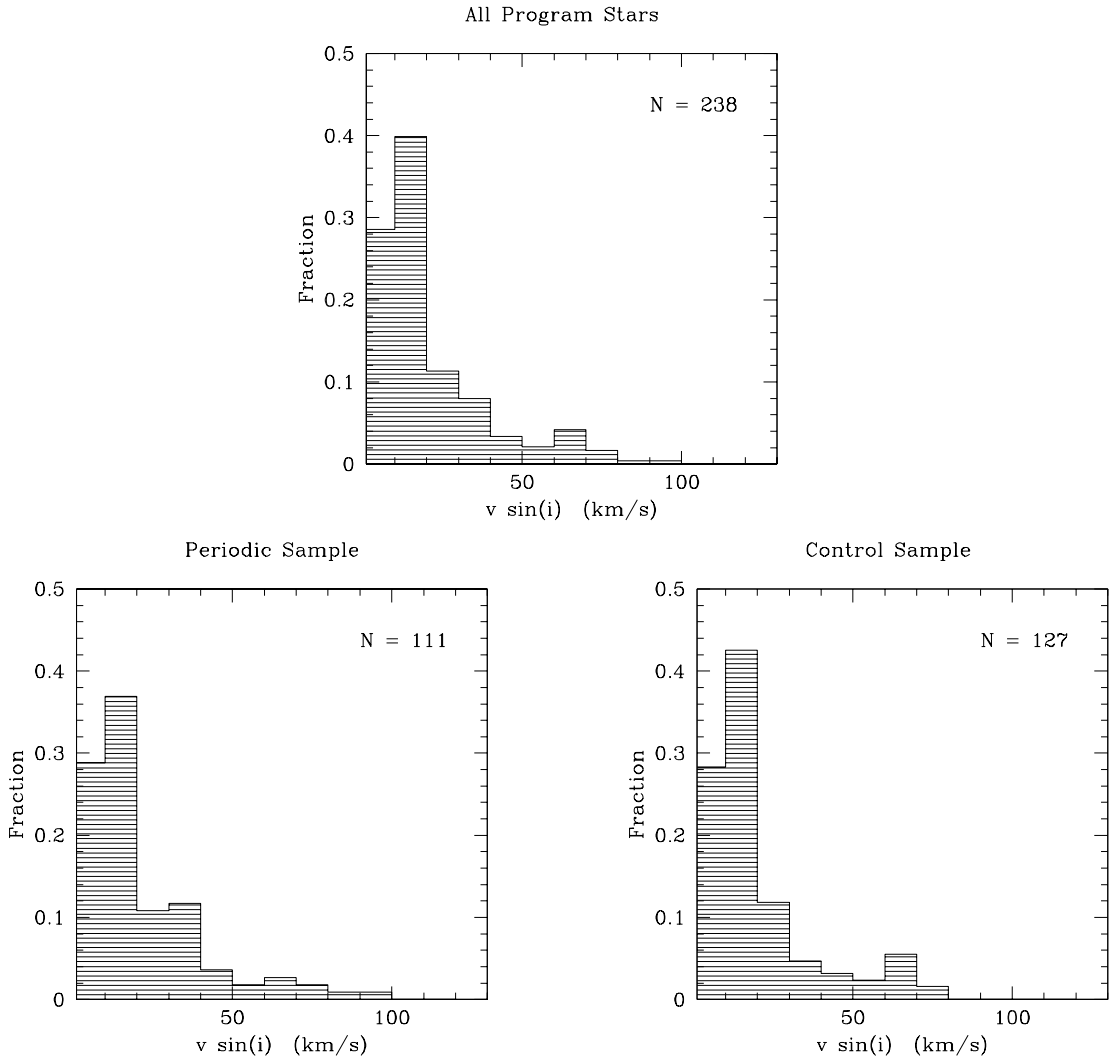}
\caption{The top plot shows the distribution of \vsini\ values for the
238 stars for which \vsini\ was successfully measured.  The bottom two
plots are the \vsini\ distributions for the Periodic and Control
sample stars.  Although there appears to be a slight excess of rapid
rotators in the Periodic sample, a K-S test comparing the \vsini\
distributions of the Periodic and Control samples shows that they are
not significantly different from each other.
\protect\label{fig:compare P and C}}
\end{figure}

The uppermost plot in Figure~\ref{fig:compare P and C} shows the
\vsini\ measurement results from this study, in the form of a
histogram of the \vsini\ distribution for all 238 stars for which a
\vsini\ was measured (including upper limits).  In this and the other
two plots shown in Figure~\ref{fig:compare P and C}, the number of
stars in each bin has been normalized by the total number of stars in
the particular sample.  Approximately 68\% of the stars have \vsini\
less than 20~\kms.  The remaining 32\% of \vsini\ values are
distributed in a ``tail'' that extends from 20~\kms\ to
$\sim$100~\kms.

One of the first systematic studies of T~Tauri star rotational
velocities was done by Vogel \& Kuhi (1981), who observed 50 stars in
the Taurus-Auriga and Orion complexes.  The Vogel \& Kuhi sample was
somewhat brighter than our sample, having $V$ magnitudes in the range
11$-$15, whereas our stars have $V$ $\sim$ 12$-$19.  Their stars were
generally more massive than ours, with masses larger than
$\sim$0.5~M$_{\sun}$.  Vogel \& Kuhi found that three-quarters of the
stars they observed had velocities below their observational limit of
25$-$35 \kms, and concluded that stars with masses $<$ 1.5~M$_{\sun}$
generally have \vsini\ less than 25 \kms.  A 1986 study by Hartmann
\et of 50 stars in Taurus-Auriga and Orion found similar results.
They observed stars with $V$ magnitudes brighter than 13$-$14, thus
comparable to the brightest stars in our sample.  Hartmann \et
concluded that nearly all the stars in the Taurus-Auriga complex have
\vsini\ $_<\atop{^\sim}$ 20\kms.  Although the large majority of our
sample stars have \vsini\ $<$ 20 \kms, 25 stars have \vsini\ $>$ 45
\kms.  Twenty of the 25 stars with large \vsini\ values have masses
less than 1.2-M$_{\sun}$, and the star with the largest \vsini\ value
is JW~526, a 0.47-M$_{\sun}$ star. Thus the finding of Vogel \& Kuhi
that stars with mass $<$ 1.5~M$_{\sun}$ have velocities less than 25
\kms\ is not true for our sample.  This is consistent with photometric
studies (e.g., Choi \& Herbst 1996, Stassun \et\ 1999, Herbst \et\
2000, 2001) which show that a number of the low-mass stars in the ONC
have short rotation periods (e.g., less than a day or two).

One of the objectives of this study was to test whether the rotation
data gained from photometric studies of pre-main-sequence stars (i.e.,
studies in which periods are measured via photometric monitoring) are
biased in some way, or are truly representative of the rotational
properties of young stars.  To address this issue, we compared the
\vsini\ distributions of the Periodic sample (the sample of stars
known to be periodic from the VVO monitoring study) and the Control
sample (the sample of stars selected to have luminosities and
effective temperatures similar to the Periodic sample stars).  (As
noted in Section~\ref{section:sample properties}, about 30\% of the
Control sample stars have since been found to be periodic.)  This
comparison is shown in the two lower plots in Figure~\ref{fig:compare
P and C}, which show the \vsini\ distributions for 111 stars in the
Periodic sample and 127 stars in the Control sample.  Both
distributions look similar to that of the full sample. One might
expect that studies based on rotation periods could be skewed toward
including a higher fraction of rapid rotators than appears in the
overall population of young stars, because rapidly-rotating stars with
more active dynamos might tend to have more prominent and/or
persistent surface spots.  In that case, the Periodic sample should
have a larger proportion of rapid rotators compared to the Control
sample.  This may at first glance appear to be true, since there is a
{\it slight} overabundance of rapid rotators in the Periodic sample
distribution.  (23\% of stars in the Periodic sample have \vsini\
$\geq$30~\kms, compared to 17\% of the Control sample stars.)
However, a Kolmogorov-Smirnov (K-S) test comparing the two samples
suggests that the difference is not significant; the test gives a
probability of 0.74, indicating that the \vsini\ distributions of the
Periodic and Control samples do not differ.  Therefore the idea that
studies of pre-main-sequence rotation are biased by the use of
photometrically-measured rotation periods is not borne out by our
data.

T~Tauri stars of similar mass and age are observed to have a wide
range of rotation periods and, in some cases, a bimodal period
distribution \citep{stassun99, herbst00}.  A proposed explanation for
the bimodality involves the concept of ``disk-locking'', in which a
magnetic interaction between the disks around T~Tauri stars and the
stars themselves regulates their rotation rates
\citep{konigl91,shu94,ostriker95}.  In this picture,
kilogauss-strength magnetic fields intercept the circumstellar disk,
channel accretion flow onto the star's surface, and ``lock'' the star
to the disk.  The rotation rate of the star is then forced to be the
same as that of the disk at the locking radius.  The specifics of
disk-locking, e.g., how long it lasts, why some stars may be locked to
their disks while others are not, and exactly what role it plays in a
star's angular momentum evolution, are unclear.  Moreover, the
observational evidence for disk-locking occurring during the T~Tauri
phase has recently been under debate \citep{stassun99,herbst00}.

In the disk-locking picture, a star that is rotating relatively slowly
is thought to be locked to its circumstellar disk (or only recently
unlocked), whereas a fast rotator either was never disk-locked or has
had time to spin up since it became unlocked.  As a test of this idea,
another useful comparison to make would be to compare the \vsini\
distributions of stars with disks and without them.  To construct a
sample of stars likely to have disks, we chose objects from our sample
with characteristics like those of Classical T Tauri Stars (CTTS).
CTTS often have strong emission lines and are thought to be
experiencing accretion onto their surfaces from a circumstellar disk,
resulting in surface hot spots that cause irregular light variations
and flaring \citep{bertout89}.  Weak T Tauri Stars (WTTS), on the
other hand, often have low-amplitude ($\leq$0.5 mag), periodic light
variations, possibly due to the presence of large cool spots on their
surfaces, and lack obvious signatures of an accretion disk
\citep{herbst94}.

We used the observational diagnostics from \citet{hillenbrand98} to
choose a sample of possible CTTS and WTTS.  These diagnostics, namely
the near-infrared excess $\Delta(I_{\rm C}-K)$ and Ca~II emission line
strength $W_{\lambda}$(Ca~II), are determined by Hillenbrand et al.\
and applied to stars in the ONC to investigate the frequency of
circumstellar disks in the cluster.  Here we adopt a combination of
the criteria used by Hillenbrand et al.  The disked (CTTS) sample we
chose consists of stars that have no measured period, and have either
(1) $\Delta(I_{\rm C}-K)$ $>$ 0.3 and Ca~II emission, or (2)
$\Delta(I_{\rm C}-K)$ $>$ 0.5.  The non-disked (WTTS) sample is
comprised of stars with measured periods, and either (1)
$\Delta(I_{\rm C}-K)$ $<$ 0.1 and $W_{\lambda}$(Ca~II) $>$ 0, or (2)
$\Delta(I_{\rm C}-K)$ $<$ 0.  We tried to use fairly stringent
criteria while still choosing a good-sized sample of objects for each
category, but the WTTS and CTTS samples are nevertheless small in
size, with 30 and 36 objects, respectively.  The \vsini\ distributions
for the WTTS and CTTS samples are shown in Figure~\ref{fig:compare
WTTS and CTTS}.  Again, the numbers in each bin are normalized by the
number of stars in the sample.  The distributions appear to differ in
the predicted sense; that is, the CTTS sample (stars likely to have
disks) does contain a slightly larger fraction of slow rotators
compared to the WTTS sample.  A K-S test comparing the two
distributions, however, yields a probability of 0.26, meaning that any
difference between them is not statistically significant.

\subsubsection{Comparing \vsini\ to Velocity Estimated from Rotation
Period}
\label{section:vsini vs. veq}

\begin{figure}
\plotone{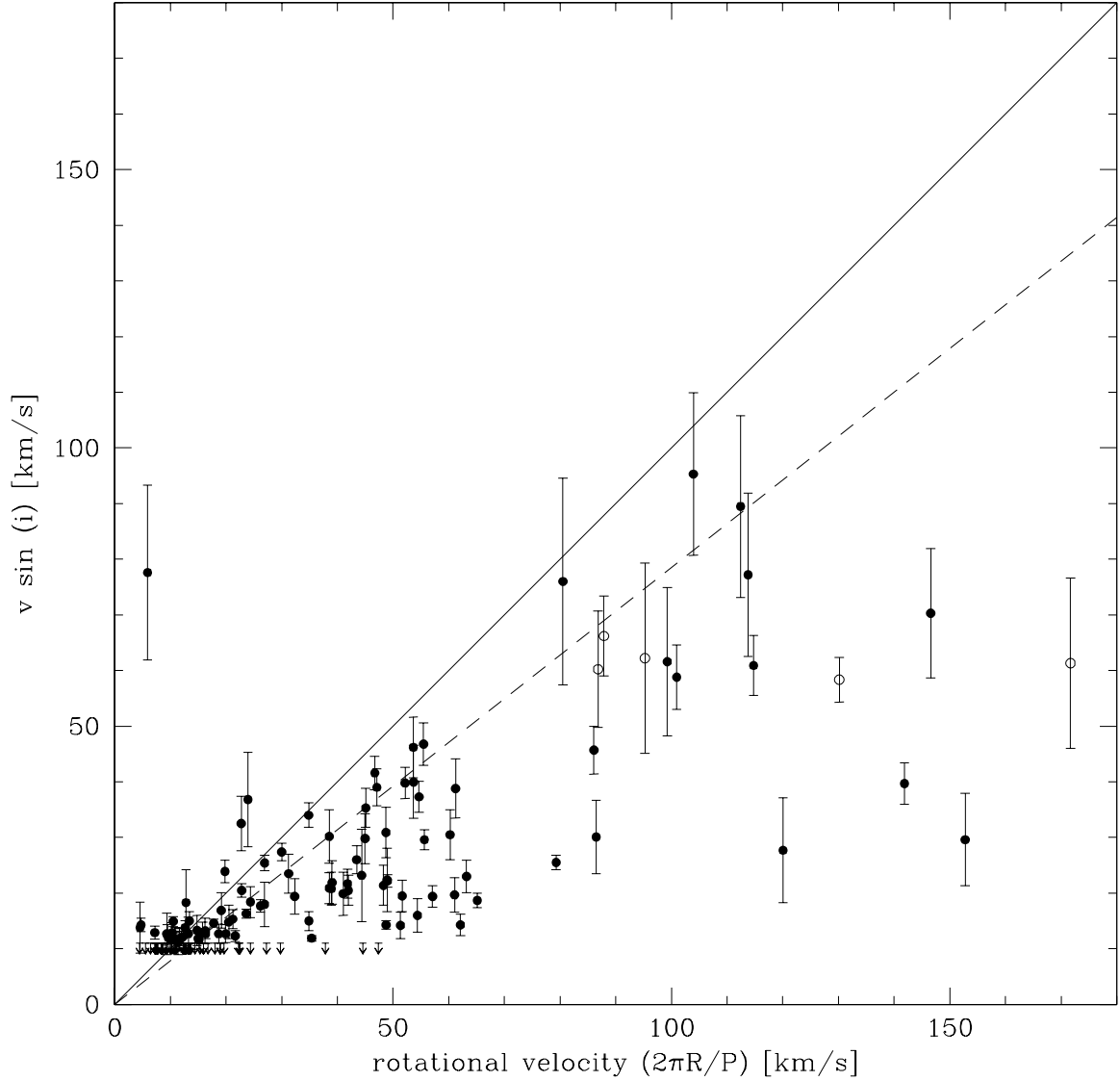}
\caption{Measured \vsini\ versus equatorial rotation velocity for the
153 stars for which the latter quantity can be calculated.  SB2
candidates are plotted with open circles, and \vsini\ values less than
or equal to 11.0~\kms\ are shown as upper limits.  Equatorial
velocities are derived from the equation $v_{eq}$ $=$ $2 \pi R / P$.
The solid line marks \vsini\ $=$ $v_{eq}$, and the dotted line marks
\vsini\ $=$ $(\pi/4)v_{eq}$, where $(\pi/4)$ is the expected average
value of $sin(i)$ for a random distribution of stellar rotation axes.
\protect\label{fig:vsini vs. veq}}
\end{figure}

Because we know the rotation periods of most of the stars for which we
measured \vsini, we can compare the equatorial rotation velocities
(calculated using the periods and radii estimated from the
luminosities and effective temperatures) for those stars with their
measured \vsini\ values.  This comparison is shown in
Figure~\ref{fig:vsini vs. veq}: \vsini\ values for 153 stars are
plotted versus equatorial velocities derived from the equation
$v_{eq}$ $=$ $2 \pi R / P$. (Note that one star --- JW~834 --- has a
\vsini\ and period, but has neither a luminosity nor an effective
temperature determination, so its radius cannot be calculated.)  SB2
candidates are shown with open circles. For stars with \vsini\ less
than or equal to our estimated limit of 11~\kms, upper limit symbols
are plotted. The solid line in the figure marks \vsini\ $=$ $v_{eq}$,
and the dotted line marks \vsini\ $=$ $(\pi/4)v_{eq}$.  Assuming
randomly-oriented stellar axes, the observed mean value of $sin(i)$
should be $\pi/4$, or 0.785 \citep{chand50}. Therefore, at least in
theory, the points are expected to scatter about the dotted line.

Figure~\ref{fig:vsini vs. veq} shows first of all that there is a
strong correlation between \vsini\ and $v_{eq}$, thus demonstrating
conclusively that the periodicity of T~Tauri stars is caused by the
rotation of stars with spotted surfaces.  The correlation of these
quantities is strong: results from two different non-parametric tests
indicate that the probability that there is no correlation is between
10$^{-24}$ and 10$^{-27}$.  In addition, the \vsini\ $=$ $v_{eq}$
limit is respected by nearly all the stars to within the errors of the
measurement.  The one notable exception, JW~275, is discussed
below. This is a gratifying result and an important confirmation of
the basic assumptions underlying pre-main-sequence rotation studies to
date.

A noticeable feature of Figure~\ref{fig:vsini vs. veq} is that the
majority of points fall below the \vsini\ $=$ $(\pi/4)v_{eq}$ line,
indicating that the average $sin(i)$ value we observe is lower than
the expected value.  To investigate the significance of this
deviation, we begin by calculating a mean $sin(i)$ including all 153
stars with both \vsini\ and $v_{eq}$ values.  This yields \msini\ =
0.82 $\pm$ 0.09, with a standard deviation of 1.1.  This result is
dominated by a few very high values of $sin(i)$ that cannot simply be
understood in terms of random errors.  The most extreme example is
JW~275, which has a radius of 2.3~R$_{\sun}$ and VVO period of 20.1
days, implying an equatorial velocity of 5.9~\kms.  Its measured
\vsini, however, is 78~\kms, completely inconsistent with expectation.
This star represents a clear example of a gross error; either its
rotation period or its \vsini\ is incorrect.  The star's period was
measured from only a single longitude, and therefore could be a beat
period.  If the measured 20.1-day period is actually the beat period
between a 1-day sampling interval and a 1.05-day rotation period, then
$v_{eq}$ for the star should be 112~\kms.  Combining this with the
measured \vsini\ yields the quite reasonable $sin(i)$ value of 0.69.
Rotation periods close to one day are difficult to distinguish from
beat periods in data obtained at a single longitude.  Clearly this
star and others like it need to be removed from the sample before
drawing inferences about the observed distribution of $sin(i)$.  Stars
whose \vsini\ measurements are too small compared to their likely
errors should also be removed before \msini\ is calculated.

Accordingly, we first restrict the sample to those stars with \vsini\
greater than 11~\kms, our approximate measurement limit based on the
discussion in Section~\ref{section:measure vel}.  We also remove from
the sample the remaining stars (which include JW~275) with predicted
$v_{eq}$ values less than 11~\kms.  The rationale for eliminating
those stars is first of all that if the period and radius are accurate
and $v_{eq}$ is less than $\sim$11~\kms, we do not expect to measure a
reliable \vsini\ given our resolution.  Alternatively, if the period
and radius are in error, the star should for that reason be culled
from the sample.  For the latter reason we exclude three stars with
$sin(i)$~$>$~1 that may be cases similar to JW~275.  Applying these
conditions leaves a sample of 86 stars with precise $sin(i)$
estimates.  This sample yields \msini\ $=$ 0.64$\pm$0.02 with standard
deviation $=$ 0.23.  (If we include the above-mentioned three stars
with $sin(i)$~$>$~1, the result is the same within the errors: \msini\
$=$ 0.67$\pm$0.03, standard deviation 0.26.)  Furthermore, if we
construct an even more robust sample of stars with \vsini\ and
$v_{eq}$ greater than 13.6~\kms, our estimate of the formal spectral
resolution, then \msini\ $=$ 0.59$\pm$0.025 (standard deviation $=$
0.21).  The average inclination for this sample of 68 objects is
38$\pm$2 degrees.\footnote{We note that the errors on the mean values
given here were calculated assuming a Gaussian distribution.  This is
not strictly valid as $sin(i)$ is not expected to be
normally-distributed.  Our actual {\it observed} $sin(i)$ values,
however, do follow a smooth, fairly symmetric distribution with a
prominent peak.  Thus we have used the mean and its standard error to
characterize the observed distribution.}

Results similar to ours have been reported in previous studies that
combined \vsini\ data with rotation periods of pre-main-sequence
stars.  \citet{hartmann86} compared \vsini\ and $v_{eq}$ for six
T~Tauri stars and found that the \vsini\ values of two of the slower
rotators in their sample fell below the expected $v_{eq}$ values.
\citet{weaver87} combined data from \citet{hartmann86} with that from
other studies to produce a sample of 20 stars for which both \vsini\
and period were known.  Weaver found that the mean inclination for the
sample of 19 stars with $sin(i)$ $<$ 1 was 40 degrees rather than the
expected value of 57 degrees.  The average $sin(i)$ value from
Weaver's data is 0.63.  More recently, \citet{soderblom99} observed 35
stars in the open cluster NGC~2264. For the $\sim$20 stars in their
sample with measured \vsini\ and period and $sin(i)$ $\leq$ 1, the
average $sin(i)$ value was 0.6, also in agreement with our result.

There are a number of possible explanations for the discrepancy
between observed average $sin(i)$ values and theoretical expectation.
To clarify the possible reasons, we express $sin(i)$ in terms of
quantities either observed directly or derived from observations by
writing:
\begin{equation}
{sin(i) = (constant)~(v~sin(i))~P~{T_{\rm{eff}}}^2~{L^{-1/2}}}
\end{equation}
Systematic errors in \vsini, luminosity, effective temperature, and/or
rotation period could give rise to the observed deviation from \msini.
The approximate size of the systematic error(s) would be 0.2/0.6, or
about 30\%.  Alternatively, the smaller-than-expected \msini\ might be
due to a real physical phenomenon.  We have explored each of these
possibilities and address them below.

{\it Physical causes.}--- There may be an actual physical cause behind
the lower average $sin(i)$ we observe.  For example, Weaver (1987)
suggested that it could be the result of selection against stars with
high axial inclination in rotation studies, because circumstellar
material obscures the stellar photosphere.  We have explored this idea
and find from numerical simulations that in order to reduce the
average value of $sin(i)$ from 0.79 to 0.60, all stars with
inclinations between 52 and 90 degrees must be systematically excluded
from period studies.  Given a random distribution of rotation axes,
this means about 60\% of the stars would have to be excluded, which
seems unlikely.  Alternatively, it is possible that rotation axes in
young clusters like the ONC and NGC~2264 are aligned rather than
oriented randomly.  Some have suggested that the formation of a star
cluster from a rotating molecular cloud threaded by a uniform magnetic
field should eventually lead to alignment of stellar rotation axes
(see Bodenheimer 1995 and references therein).  Evidence with regard
to such alignment is mixed, with some data showing alignment of
structures within molecular clouds (see Heiles \et 1993 and included
references) and some indicating that stellar rotation axes are
randomly oriented in certain clusters (e.g., the Pleiades; Stauffer
1991).  Another explanation could be that the prominent surface spots
that give rise to T~Tauri stars' photometric variability are more
likely to appear near the poles of the stellar rotation axis than near
the equator.  Consequently the stars that are detected in photometric
period studies (and thus appear in Figure~\ref{fig:vsini vs. veq})
would tend to have more pole-on inclinations, and thus have $sin(i)$
values that are truly lower than expected from a random distribution.
On the other hand, if a star were seen exactly pole-on, there would be
no photometric variation observed.  In any case,
Figure~\ref{fig:compare P and C} shows that there is not a significant
bias between the \vsini\ distributions of stars selected from period
studies (the Periodic sample) compared to those selected via only
their luminosities and effective temperatures (the Control sample).

{\it Systematic errors in \vsini.}--- One way to produce a
smaller-than-expected average $sin(i)$ value is if our measured
\vsini\ values are systematically low.  Increasing our values by at
least 30\% would bring \msini\ close to its expected value.  The
comparison of our \vsini\ values for the 14 stars we have in common
with other studies (see Figure~\ref{fig:compare vsini}) shows,
however, that for over half the stars, our measured \vsini\ values are
actually slightly larger than the values from other work.  The overall
mean difference between our \vsini\ values and those of the two other
studies is 0.5$\pm$2.0~\kms.  This, coupled with the fact that other
independent studies of pre-main-sequence stars in NGC~2264, Taurus and
Orion also found average $sin(i)$ values of approximately 0.6, makes
it seem unlikely that a systematic \vsini\ error is wholly responsible
for the \msini\ discrepancy.  On the other hand, it is difficult to
rule out the possibility that some smaller systematic error in \vsini\
is contributing to the problem.  In Section~\ref{section:vsinis} we
explored systematic uncertainties associated with line blending and
possible mismatches in spectral type between the targets and
cross-correlation templates.  In those cases the associated
uncertainties were accounted for by our estimated errors.  But there
could be other systematic effects at work.  For example, narrow-lined
spectra from dwarf stars are used as cross-correlation templates
against which the pre-main-sequence stellar spectra are compared.  It
is feasible that a difference in the limb darkening law between dwarfs
and pre-main-sequence stars could have an effect on the \vsini\
calibration, introducing a systematic error.  Such a systematic error
could potentially affect the other \vsini\ studies mentioned, most of
which also used dwarfs as templates for the cross-correlation.

{\it Errors in the photometric periods.}--- Still another explanation
for the lower average $sin(i)$ value could be that the rotation
periods inferred from photometric studies are, in some cases, not the
actual rotation periods of the stars.  Instead, they could be false
detections, harmonics of the rotation period, beat periods between an
observing interval (normally one day) and the rotation period, or, in
the case of spectroscopic binaries, related instead to the orbital
period of the system.  We consider each of these ideas in turn.

It is unlikely that many of the photometric periods are false
detections rather than true periods, since statistical tests are used
to eliminate these and, in many individual cases of low $sin(i)$
values, there are confirming period determinations at different epochs
(see Herbst et al. 2000). Moreover, the presence of false detections
should produce a random scatter in the \vsini$-$$v_{eq}$ plane, tending
to obscure the observed correlation rather than skew it towards small
values of $sin(i)$.  Harmonics, on the other hand, could explain at
least some of what is displayed in Figure~\ref{fig:vsini vs. veq}.
T~Tauri stars such as V410~Tau \citep{herbst89} and some stars in the
ONC \citep{stassun99, herbst01} have displayed so-called ``period
doubling'', i.e., the period determined at one epoch is later shown to
be half of the true rotation period because the star originally had
significant spots on two opposing hemispheres.  The effect of
correcting a period-doubling error would be to decrease $v_{eq}$ by a
factor of two for any such star.  Taking the 68-star sample described
above and arbitrarily doubling the period of every star with an
inferred $sin(i)$ $<$ 0.60 (34 of the 68 stars) increases \msini\ to
the expected value.  Whether this is part or all of the problem can
only be determined by continued careful monitoring of ONC stars to see
whether period doubling occurs for a significant number of them.
Comparisons made so far indicate that this is {\it not} a major
contributor to the problem (Herbst \et\ 2001; Carpenter, Hillenbrand
\& Skrutskie 2001).

Another explanation for those stars with larger-than-expected $v_{eq}$
could be that the observed photometric periods are actually beat
periods.  This would preferentially affect stars with shorter periods
(i.e., $\sim$1 day or so) since the periods are closer to the
observing frequency.  Both the VVO \citep{herbst00} and ESO
\citep{herbst01} data sets were obtained at single longitudes
(although with multiple observations per night) and are thus more
susceptible to the beat phenomenon than the \citet{stassun99} study,
which combined data from widely-separated longitudes.
\citet{herbst01} compared their data with \citet{stassun99} and found
that of 111 stars monitored by both studies, 98 yielded identical
periods, two were likely examples of period doubling, one appeared to
be a false detection, and the other 10 were related as beat periods
with a 1-day sampling interval.  This suggests that about 10\% of the
sample of periodic stars might be contaminated by beat periods or
harmonics.  Again, the continued monitoring of ONC stars, preferably
from observatories well separated in longitude, is necessary in order
to determine whether these phenomena are important in the
interpretation of Figure~\ref{fig:vsini vs. veq}.

Lastly, it is possible that some of the periodicity detected in
photometric studies is of orbital origin.  Close spectroscopic
binaries are variable for a number of reasons including eclipses,
tidal distortion and the reflection effect.  Such stars may not be
rotationally locked and observed periods could be the orbital period
or a related harmonic.  Alternatively, in the case of tidal locking,
both tidal distortion and reflection can lead to period doubling.  As
mentioned in Section~\ref{section:vsinis}, seven stars show evidence
indicating that they might be spectroscopic binaries.  Five of these
stars have measured periods and are marked with open circles in
Figure~\ref{fig:vsini vs. veq}.  Of the five stars with the largest
$v_{eq}$ values in the figure, two are SB2 candidates.  If these
stars' observed periods arise from orbital motion rather than
rotation, then they should not be plotted on Figure~\ref{fig:vsini
vs. veq}.

In summary, despite the fact that the rotation periods are more
precise than the other quantities contributing to $sin(i)$ (typically
being known to $\sim$1\% or better; see, e.g., Choi \& Herbst 1996),
they are prone to gross errors such as aliasing and period doubling.
These effects probably account for the small number of outliers
removed from the sample used to calculate our observed \msini.  It is
difficult to see how they could account for the remaining systematic
effect, which is accompanied by a relatively small scatter in
$sin(i)$.  In order for such period errors to account for the
remaining systematic effect, a significant number of rotation periods
would need to have been underestimated.  This could be the case if
period-doubling plays a major role.  This seems unlikely, however, in
light of the general agreement found between periods measured by
different observers at different epochs with different sampling
intervals (e.g., Stassun et al. 1999, Herbst et al. 2001, Rebull 2001,
Carpenter \et\ 2001).

{\it Overestimated luminosities.}--- Equation (1) shows that $sin(i)$
depends on $L^{-1/2}$; as a result, systematically overestimated
luminosities for our target stars could give rise to a
lower-than-expected \msini.  For the sample of 68 stars, the true
stellar luminosities would have to be approximately 60\% of the
current measured values in order to yield an average $sin(i)$
consistent with the expected value.  Weaver (1987) has suggested that
the observed luminosities of pre-main-sequence stars might be
artificially inflated due to the presence of a circumstellar disk.
This idea could explain why the $sin(i)$ discrepancy apparent in our
data has been seen by others (namely Soderblom et al. 1999 and Weaver
1987), who have observed pre-main-sequence stars in different clusters
with different observational methods.  On the other hand, this
requires that the large majority of our target stars have a relatively
luminous disk, i.e., one that is responsible for $\sim$40\% of the
total optical luminosity of the disk-star system.  Hillenbrand \et\
(1998) estimate the disk fraction in the ONC to be between 55\% and
90\%, so the majority of ONC stars may indeed have disks.  However
when we examine the $\Delta(I_{\rm C}-K)$ values of the 68-star
sample, we find that only about 40\% have $\Delta(I_{\rm C}-K)$ $>$
0.3, indicating the presence of a prominent disk.  There are
uncertainties associated with determining $\Delta(I_{\rm C}-K)$, and
Hillenbrand \et\ note that the errors (which are expected to be
primarily negative, reducing $\Delta(I_{\rm C}-K)$ from its true
value) may be as large as 0.1$-$0.3 magnitudes for stars with spectral
types K2 to M3.  Nevertheless it seems unlikely that the large
majority of our stars have disks that are sufficiently luminous to
account for the observed \msini\ discrepancy.

Another uncertainty associated with the luminosities is the extinction
correction used to derive them.  Systematically overestimated
extinction values would result in too-large luminosities.  Our
luminosities come from Hillenbrand (1997), who suggests that if
anything the luminosity values may be {\it underestimated}, if there
is unaccounted-for accretion activity or a nonstandard or
spatially-variable extinction law in the ONC.  In calculating the
overall uncertainty on the luminosities, Hillenbrand assumes an error
in A$_{\rm{V}}$ of 0.5 magnitudes, which is larger than expected and
which corresponds to an uncertainty of 0.12 in log($L_{*}/L_{\sun}$).
Other sources of error that may contribute are intrinsic uncertainties
due to photometric variability, errors from translating spectral type
and color to luminosity, and uncertainties in the stellar distances.
Hillenbrand estimates that these contribute uncertainties of 0.12,
0.15 and 0.075 dex, respectively, for a total combined uncertainty of
less than 0.2 dex, on average.  Subtracting this amount from the
luminosities for the sample of 68 stars yields \msini\ $=$
0.75$\pm$0.03, which is close to the expected value.  But again,
Hillenbrand suggests that if a systematic error is present in the
luminosities, it would likely go the other way, leading to
underestimated values instead of too-large ones.

A detailed discussion of the observational uncertainties associated
with estimates of stellar luminosities is given in \citet{hartmann01},
who explores the effect of such errors on age estimates in
star-forming regions.  In addition to discussing each of the sources
of error mentioned above, Hartmann suggests that unresolved binary
companions can have a potentially large effect on luminosity
determinations.  Moreover, this would be a systematic effect: the
presence of unidentified companion stars around our targets would bias
their luminosities to higher values.  Hartmann models the effect of
unresolved binaries on luminosities and estimates that the shift in
log~$L$ would be $\sim$0.2, under the assumption that nearly all stars
have unidentified companions.  As explained above, a shift of this
magnitude could bring \msini\ close to its expectation value, so this
idea could potentially help explain the observed discrepancy.

{\it Underestimated effective temperatures.} Because $sin(i)$
$\propto$ $T_{\rm{eff}}^{2}$, the $sin(i)$ values we derive are much
more sensitive to errors in the effective temperatures than to errors
in the luminosities.  The effective temperatures we use come from
Hillenbrand (1997), who determined spectral types for ONC stars and
then converted them to $T_{\rm{eff}}$ values using the calibration of
\citet{ck79}, with a few modifications.  We find that increasing
$T_{\rm{eff}}$ for our sample stars by 400 to 600 Kelvin (without
changing the luminosity) yields good agreement between the measured
and expected $sin(i)$ distributions.  Figure~\ref{fig:vsini vs. veq
newT} shows the effect of a 600-K increase in temperature in the
\vsini\ versus $v_{eq}$ plane.  All 153 stars that have \vsini\ and
$v_{eq}$ measures are shown.  The points now scatter around the dotted
line representing \vsini\ $=$ ($\pi$/4)$v_{eq}$.  The figure
demonstrates that increasing the temperatures can rectify the \msini\
discrepancy; however, this fix of the problem requires that the
temperatures have been significantly underestimated.

An examination of the literature suggests that it is possible that
uncertainty inherent in the conversion from spectral type to
$T_{\rm{eff}}$ could cause the temperatures to be off by the required
amount.  \citet{padgett96} and \citet{king98}, for example, discuss
problems with the existing spectral type$-$$T_{\rm{eff}}$ calibration
for pre-main-sequence stars. These include the uncertainty with regard
to whether a dwarf or giant temperature scale is more appropriate, as
discussed by \citet{hillenbrand97} and \citet{white99}.  King
estimates that the size of the errors in $T_{\rm{eff}}$ determinations
for pre-main-sequence stars is 500$-$700~K.  Padgett finds differences
as large as 960~K between $T_{\rm{eff}}$ values derived from line
ratio measurements in high resolution spectra and those obtained from
spectral~type-$T_{\rm{eff}}$ calibration.  She finds that temperatures
derived from metallic line ratios are generally higher than those
determined from lower-resolution spectral classification.  The
assumption that pre-main-sequence stellar atmospheres can be precisely
calibrated by comparison with the atmospheres of dwarfs or evolved
stars is likely to break down at some point.  A major difference is
that pre-main-sequence stars have spots --- indicating the presence of
strong magnetic fields --- covering a large portion of their surfaces.
The effect of these spots on the spectral type-$T_{\rm{eff}}$
calibration is not well-understood.  Moreover, \citet{ck79} note that
it is unclear exactly what the physical meaning of $T_{\rm{eff}}$ is
in a star with different emitting regions (such as the photosphere,
chromosphere, extended atmosphere, and circumstellar disk) that each
contribute significant amounts of flux.

To summarize, we find a significant departure of the distribution of
measured $sin(i)$ values from what is expected for a random
distribution of observed inclination angles.  This departure has also
been seen in previous studies of pre-main-sequence stars.  It could
have a number of explanations, including real physical conditions,
selection effects, and/or systematic errors in one or more of the
quantities that go into calculating $sin(i)$.  If this were entirely
due to an error in the spectral type-to-$T_{\rm{eff}}$ calibration, an
adjustment by 400 to 600 Kelvin towards hotter temperatures than
determined by Hillenbrand (1997) is required to solve the
problem. This is large, but perhaps not impossible, given the current
state of our understanding of effective temperatures of
pre-main-sequence stars.  If the $T_{\rm{eff}}$ values of the stars
have been underestimated, the masses and ages (determined by plotting
the stars on an H-R diagram and interpolating the masses and ages from
evolutionary model tracks) also need to be adjusted.  (For example, a
0.25-M$_{\sun}$ star becomes a 0.4-M$_{\sun}$ star if its temperature
is increased by 600 K.)  This in turn would have important
implications for conclusions drawn about pre-main-sequence stellar
evolution in the ONC and other clusters and associations.

\subsection{Stellar Radii}
\label{section:radii}

A main goal of this study is to combine the \vsini\ measurements we
obtain with rotation period measurements from other studies in order
to constrain the radii of the stars in our sample.  Because
$P$~\vsini\ $=$ $2~\pi~R~sin(i)$, the \vsini\ and period of a star can
be used to calculate its minimum radius, $R~sin(i)$.  With enough
data, the rotational velocities and periods can provide statistical
estimates of radii (without the $sin(i)$ ambiguity) for stars 
grouped by their similar characteristics.  Direct measurements of the
radii of pre-main-sequence stars are hard to come by, since so far
only four pre-main-sequence eclipsing binaries have been discovered
\citep{popper87,casey95,casey98,andersen91,mamajek00,covino00}.  Therefore
other methods for independently estimating stellar radii can be
valuable tests of the reliability of the radii of pre-main-sequence
stars and hence the accuracy with which they can be placed on the
theoretical H-R diagram.

\begin{figure}
\plotone{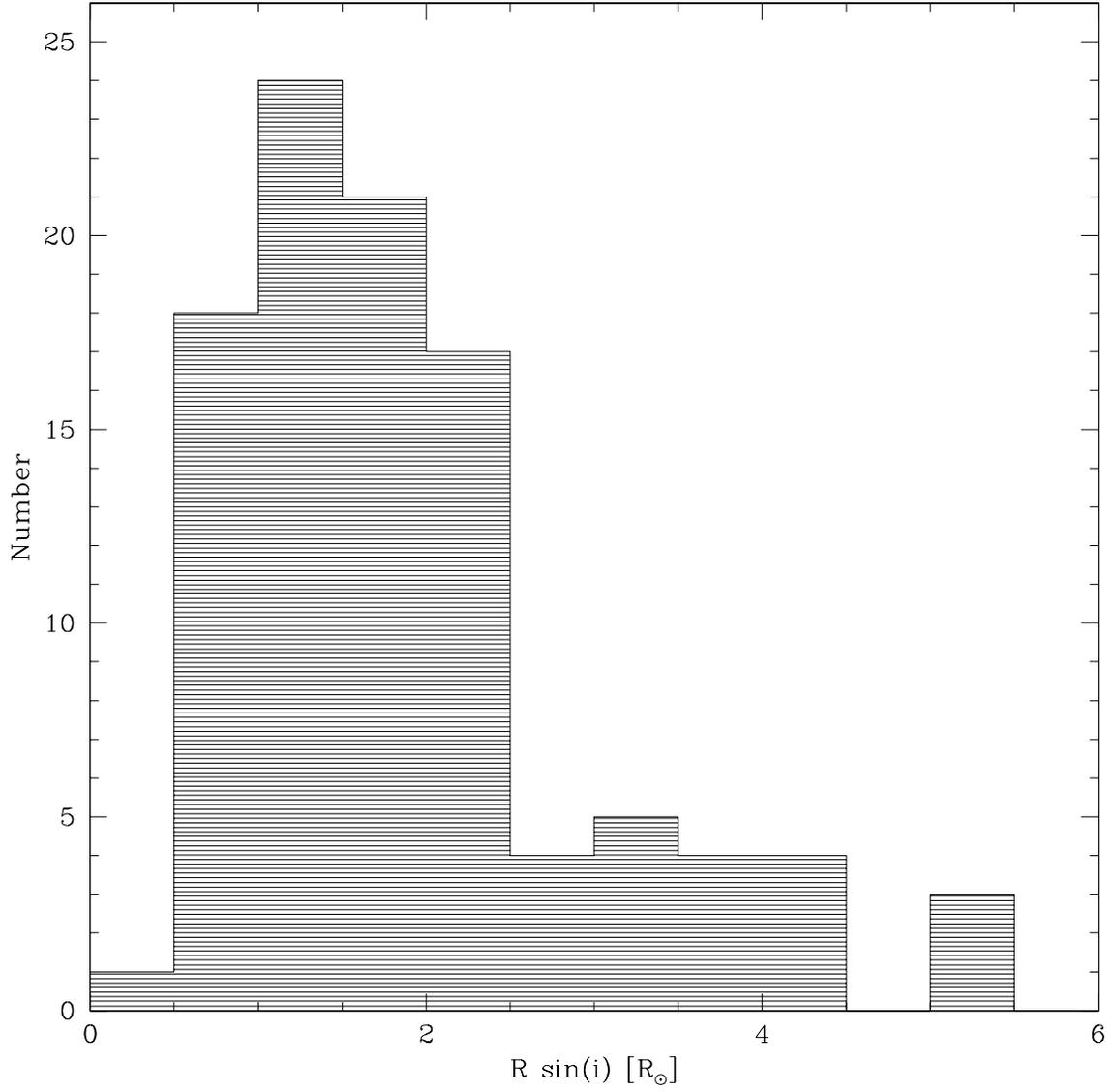}
\caption{Minimum radii for 101 stars with measured period and
\vsini\ $>$ 11.0~\kms, calculated from the equation $R~sin(i)$ $=$
$P$~\vsini$/(2\pi)$.  80\% of the stars have $R~sin(i)$ $<$
2.5~R$_{\sun}$.  \protect\label{fig:rsini}}
\end{figure}

We have combined period and \vsini\ measurements to calculate minimum
radii for individual stars with \vsini\ greater than our estimated
limit of 11~\kms.  It should be emphasized that these minimum radii
are not subject to the uncertainties in luminosity and effective
temperature discussed in the previous section, but instead rely {\it
only} on the measured period and \vsini.  There are 101 such stars not
including JW~275, which has a period that is likely in error, as noted
in Section~\ref{section:vsini vs. veq}.  (Calculating a minimum radius
using JW~275's measured period of 20.1 days yields an $R~sin(i)$ of
30.8~R$_{\sun}$; if its true period is instead 1.05 days, its
$R~sin(i)$ is 1.61~R$_{\sun}$.)  Minimum radii for the 101 stars are
given in Table~\ref{table:all data} in units of the solar radius, and
Figure~\ref{fig:rsini} shows a histogram of the values.  They range
from less than one to more than five solar radii, with the majority
(80\%) having $R~sin(i)$ $<$ 2.5~R$_{\sun}$.  The median of the
distribution is 1.7~R$_{\sun}$.

Next we turn to making statistical estimates of the radii of groups of
stars located in the same part of the H-R diagram.  This calculation
requires a sample of stars with reliable \vsini\ and period
measurements, as well as known $L$ and $T_{\rm{eff}}$ values.  $L$ and
$T_{\rm{eff}}$ are needed only so that the stars can be grouped in a
way that will make the radii derivation meaningful; again, {\it the
radii we calculate are completely independent of these two
quantities.}  The same potential sources of error that were relevant
when we explored the magnitude of the \msini\ discrepancy in
Section~\ref{section:vsini vs. veq} must be considered here as well.
For example, stars with $sin(i)$ values much larger than 1 may have
rotation periods that are in error.  In addition, \vsini\ values near
our estimated measurement limit (11~\kms) may be less certain than
larger values.  To derive the most robust average radii possible with
these data, we use the 68-star sample --- consisting of stars with
accurate periods and \vsini\ $\geq$13.6~\kms\ --- that was constructed
to help characterize the \msini\ problem.

\begin{figure}
\plotone{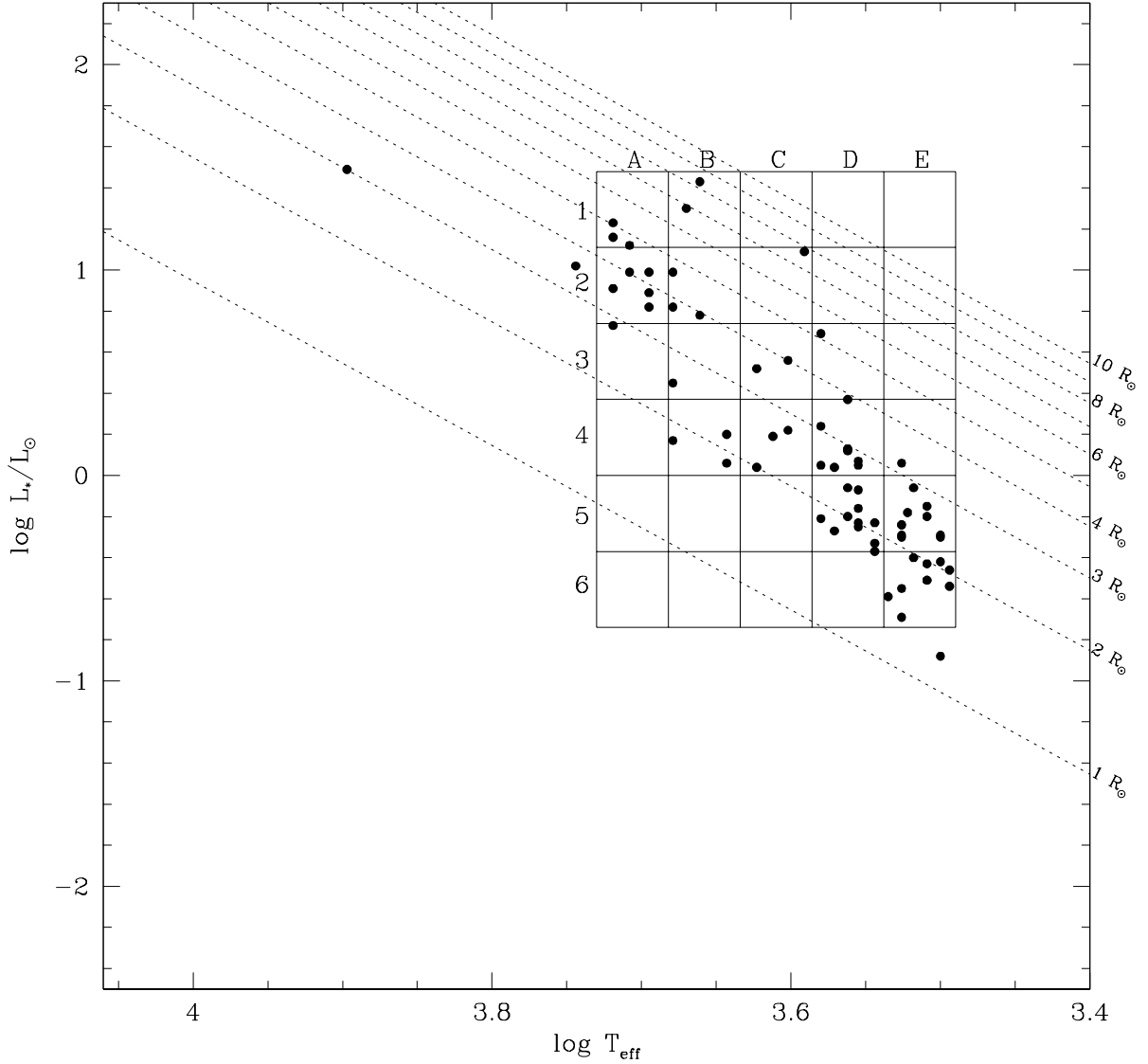}
\caption{This figure shows the locations of the radii bins described
in Section~\ref{section:radii}.  The bins are marked with solid lines;
each is 0.048 units wide in log~$T_{\rm{eff}}$ and 0.37 units wide in
log~$(L/L_\sun)$.  The bin numbers are designated in the figure; e.g.,
the bin in the bottom right corner is bin E6.  68 stars with \vsini\
$\geq$ 13.6~\kms\ and known rotation period, luminosity and effective
temperature appear as filled circles.  Dotted lines mark lines of
constant radii in the log-$L$, log-$T_{\rm{eff}}$ plane, ranging from
1$-$10 R$_{\sun}$ in increments of 1.0. For bins containing two or
more stars, average radii have been calculated and are given in
Table~\ref{table:radii}.  \protect\label{fig:radii bins}}
\end{figure}

We want to calculate average radii for groups of stars that are in
approximately the same evolutionary stage.  Consequently the stars
must be binned according to their location in the theoretical H-R
diagram.  Even if there is a systematic error in $T_{\rm{eff}}$ such
as that discussed in Section~\ref{section:vsini vs. veq}, the {\it
relative} locations of the sample stars should be fairly constant, so
that grouping them by effective temperature should still be
reasonable.  Using as a guide the way in which the stars are arranged
on the H-R diagram, a series of equal-sized bins was chosen that would
maximize the number of stars within each bin (and the number of stars
that could be binned) while still retaining adequate resolution to
derive meaningful average radii.  The bins are shown as solid lines in
Figure~\ref{fig:radii bins}.  They each cover 0.048 units in
$log~T_{\rm{eff}}$ space and 0.37 units in $log~(L/L_\sun)$ space.
The 68 sample stars used here are shown as filled circles; three of
them fall outside of the chosen bins.  The dotted lines marked on the
figure are lines of constant radius, at intervals of 1~R$_\sun$,
beginning at the bottom with 1~R$_\sun$.

Eleven of the 30 bins contain at least two stars. For these bins, an
average stellar radius was calculated, by first computing $R~sin(i)$
for each individual star using the measured \vsini\ and period,
averaging the projected radii for the stars in the bin, and finally
dividing the average $R~sin(i)$ by the average expected value of
$sin(i)$, $\pi/4$.  The radii are given in Table~\ref{table:radii}.
The first three columns of the table specify the bin number and
luminosity and effective temperature range spanned by the bin.  Column
4 gives the number of stars in the bin.  Column 5 gives the mean
$R~sin(i)$ value and its error, and column 6 lists the mean radius and
error.  The last column lists, for comparison, the average radius
corresponding to the center of each $L$ and $T_{\rm{eff}}$ bin.  The
mean difference between the radii calculated from \vsini\ and $P$ and
the radii for the centers of the bins is $-$0.70, with a standard
deviation of 0.74. Some difference between the $R$ values calculated
using \vsini\ and period and those based on the locations of the stars
in the H-R diagram is of course expected, in light of the data shown
in the \vsini\ versus $v_{eq}$ plot and the accompanying discussion.

The calculation of $\langle$$R$$\rangle$ from period and \vsini\
yields a measurement of the radii of our sample stars that is
independent of the usual methods for deriving stellar radii, and
provides a snapshot of the evolution of low-mass stars as they
contract toward the main sequence.  To investigate whether we can
actually observe that stars are contracting (i.e., use our data to
show this geometrically), we have grouped the bins in
Table~\ref{table:radii} according to their $T_{\rm{eff}}$ values, then
ordered bins with the same $T_{\rm{eff}}$ values by decreasing
luminosity.  Figure~\ref{fig:hrd} shows that the evolutionary tracks
in the portion of the H-R diagram in which our stars are located run
fairly vertically.  Therefore, examining the average $R~sin(i)$ values
in sets of vertical bins (i.e., comparing bin A1 to A2, B1 to B2,
etc.)  should tell us whether stars of similar mass are actually
contracting as they approach the Zero Age Main Sequence (ZAMS).

Examination of Table~\ref{table:radii} shows that the mean radii do
decrease with decreasing luminosity for bins in the same temperature
range.  However, the relative errors on \mrsini\ are as large as
20$-$30\% in some cases.  As a result, there is evidence for
contraction at greater than the three-sigma level in only one pair of
bins, B2 to B4.  This is the only instance in which non-adjacent bins
can be compared to each other.  So although the mean values do
decrease, we cannot state with absolute certainty that we are
observing pre-main-sequence contraction {\it for stars within a given
temperature range}.  If we could include more stars with reliable
\vsini\ and period in each bin (and by doing so reduce the errors on
\mrsini) or could compare bins that span a larger range in radius, we
might find more definitive evidence for contraction of stars in a
specific temperature range.

We can look for overall evidence of contraction for the {\it entire
sample} of 68 stars by combining the information in the five sets of
vertical bins shown in Table~\ref{table:radii}.  For each set of
vertical bins (A, B, C, D, and E) listed in the table, we can
calculate the relative decrease in $\langle$$R~sin(i)$$\rangle$ with
decreasing luminosity.  For example, the relative change in radius
($\Delta~R~sin(i)/R~sin(i)$) for bins A1 to A2 is 0.21.  Calculating
this quantity for each of the five sets of vertical bins and averaging
the result yields a mean relative change in projected radius of
0.49$\pm$0.16.  This result is positive --- i.e., shows that stars in
our sample are contracting as they approach the main sequence --- at
the three-sigma level.  This is the first time that pre-main-sequence
contraction has been verified geometrically.

To conclude, these data show marginally significant evidence
supporting a real spread in radius (and therefore age) at a given mass
in the ONC. It is clear that to proceed further (for example,
determining an accurate age spread for the cluster) requires a larger
sample of stars with accurate period and \vsini\ measurements.
However, the promise of this method has now been clearly demonstrated.

\section{Summary \& Conclusions}
\label{section:conclusions}

We have obtained high-dispersion spectroscopic data for 256
pre-main-sequence stars in the Orion Nebula Cluster, with the aim of
measuring their \vsini\ values.  Our intention was to combine the data
with that from previous studies of the ONC that measured rotation
periods via photometric monitoring, in order to better understand the
rotational evolution of T~Tauri stars and to arrive at statistical
estimates of the sizes of these stars.  A set of 118 objects with
known periods (the Periodic sample) was chosen along with 138 stars
from the same portion of the H-R diagram whose periods were not known
(the Control sample).  Most of the targets are low-mass ONC members
with spectral types from early-K to mid-M.  Spectra with sufficient
signal-to-noise to measure \vsini\ were obtained for 238 of the 256
targets.  Our results are as follows:

\noindent 1. Approximately two-thirds of the 238 sample stars with
good-quality spectra have \vsini\ less than 20~\kms, and the \vsini\
values of the remaining stars are distributed in a tail extending to
$\sim$100~\kms.

\noindent 2.  We have compared the \vsini\ distributions of the
Periodic and Control samples in order to determine whether photometric
rotation period studies are truly representative of the rotational
properties of pre-main-sequence stars.  The differences between the
distributions are not statistically significant.  The same is true
when we compare a sample of stars likely to be accreting material from
their circumstellar disks (CTTS), and perhaps ``disk-locked'', to a
sample that does not show evidence for accretion (WTTS).

\noindent 3.  We find a strong correlation between \vsini\ and
equatorial rotation velocity $v_{eq}$ derived by combining the stars'
periods with radii calculated from their published $L$ and
$T_{\rm{eff}}$ values.  This shows definitively that the observed
periodicity of T~Tauri stars is, in the majority of cases, caused by
the rotation of spots on their surfaces.

\noindent 4. The average $sin(i)$ value for our stars, calculated from
$sin(i)$ $=$ \vsini/$v_{eq}$, is significantly lower than the mean
value expected for a random distribution of stellar rotation axes.
This result, which has been seen in previous studies with smaller samples,
could be due to systematic errors in one or more of the observed
quantities that go into the $sin(i)$ calculation (namely \vsini,
period, $L$, and $T_{\rm{eff}}$), or to a real physical phenomenon.
We explore these possibilities and find that the correct $sin(i)$
value is produced if we assume that the effective temperatures of our
program stars have been underestimated by 400$-$600 Kelvin.

\noindent 5. We have calculated minimum radii for stars that have
rotation period measurements and \vsini\ greater than our estimated
measurement limit.  The $R~sin(i)$ values range from 0.4 to
5.3~R$_{\sun}$, with four-fifths of the sample falling between 0.5 and
2.5~R$_{\sun}$.

\noindent 6. We have derived average radii for groups of stars with
similar masses (i.e., in similar locations on the H-R diagram).  The
mean radii for stars within the same mass/temperature range do
decrease as the stars get closer to the ZAMS, but in most cases the
uncertainties in the mean radii are too large to show definitively
that we have observed contraction within a given mass range.  However,
we find that the mean relative change in radius with luminosity for
the five different mass/temperature ranges is 0.49$\pm$0.16, meaning
that we do find evidence for pre-main-sequence contraction in the
overall sample of stars used at the three-sigma level.

In conclusion, combining \vsini\ measurements with photometric periods
can yield valuable insight into the astrophysics of T~Tauri stars.
The methods used here allow us to test the luminosities and effective
temperatures of T~Tauri stars and thereby test their masses and ages,
since these are usually derived by comparing $L$ and $T_{\rm{eff}}$
with pre-main-sequence evolutionary models. Furthermore, even with a
modest number of stars with good-quality \vsini\ values and periods,
we have shown geometrically for the first time that the stars in our
sample are undergoing contraction as they evolve toward the main
sequence.  Applying the method to even larger samples of stars with
well-determined \vsini's and periods is likely to improve our
understanding of this important stage of low-mass stellar evolution.

% Authors may indicate to the editorial staff where they would like 
% figures and tables to be placed in the manuscript.  This is done with
% either the \placefigure{KEY} or \placetable{KEY} commands.  These
% commands require \label{KEY} commands to be placed appropriately with
% corresponding table and figure captions.  When the manuscript is
% printed a short note is printed on the page where the figure or table
% is to go.  These commands are ignored in the aaspp4 and aas2pp4 styles.

%\placetable{tbl-3}
%\placefigure{fig1}

\begin{figure}
\plotone{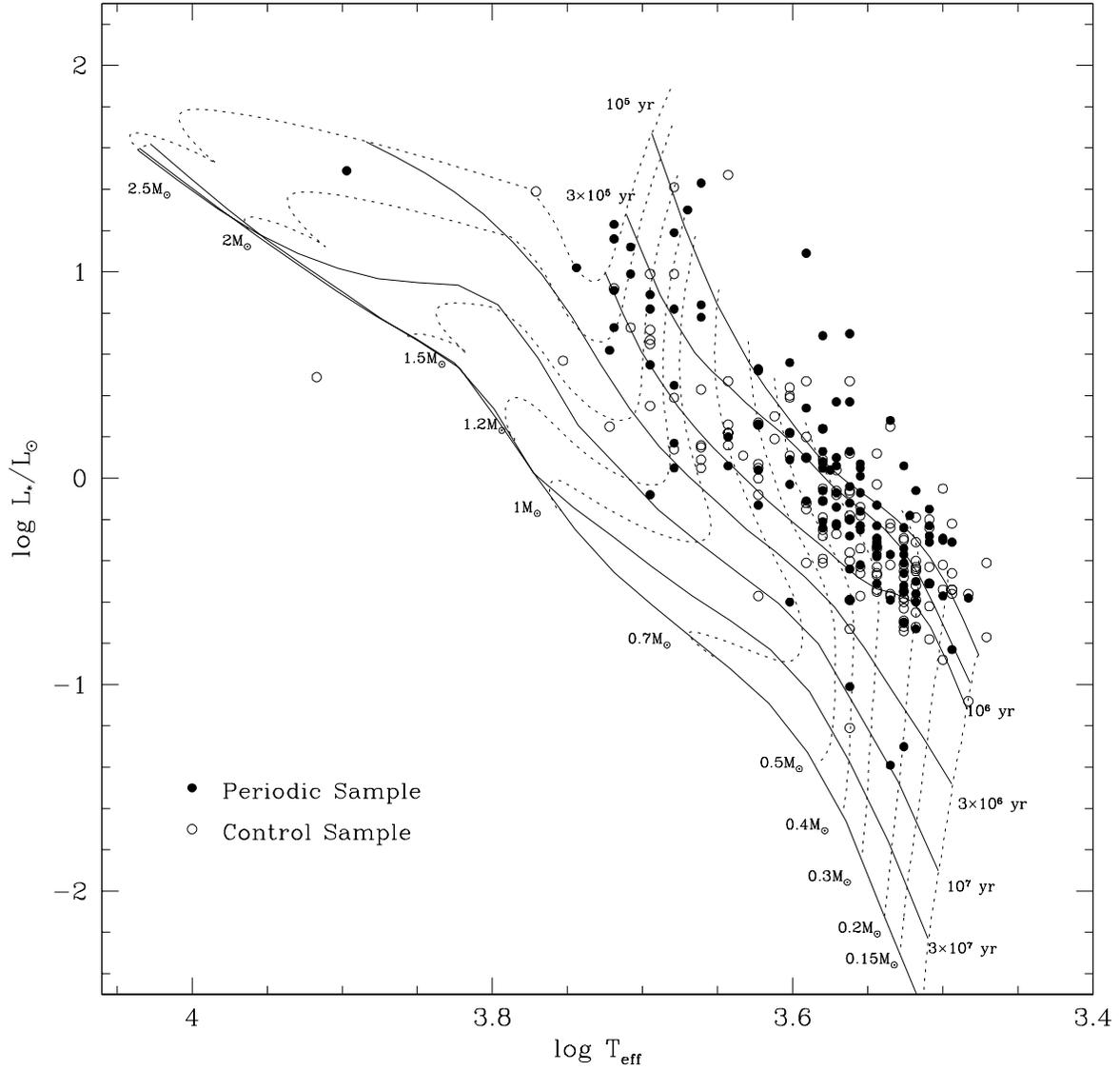}
\caption{The positions on the H-R diagram of the 256 stars included in
this study are shown here.  Filled circles are stars in the Periodic
sample, and open circles are Control sample stars.  Luminosity and
effective temperature values are from Hillenbrand (1997).
Pre-main-sequence evolutionary tracks are from D'Antona \& Mazzitelli
(1994, model 1).  \protect\label{fig:hrd}}
\end{figure}

\acknowledgments K.L.R. gratefully acknowledges a grant from the Sigma
Xi Scientific Research Society, which provided funding for the first
observing run.  We are grateful also to Lynne Hillenbrand for
furnishing the auxiliary data so essential to the design and analysis
of the project.  We have benefited from useful discussions with Lynne
Hillenbrand, Debra Fischer, and C\'esar Brice\~no.  Chris Dolan
provided advice and assistance with the spectral data reduction during
the initial phase of the project.  We would like to thank WIYN queue
observers Di Harmer and Daryl Willmarth for executing the December
1997 observations.  The H-R diagram with pre-main-sequence tracks
shown in Figure~\ref{fig:hrd} was made using plotting macros written
by Karen Strom.  Finally, we thank John Salzer, C\'esar Brice\~no,
Lynne Hillenbrand, and the referee Steve Strom for providing useful
comments and suggestions that improved the paper.  This work was
supported in part by a NASA grant through its Origins program to
W.H. and by NSF grant AST-9417195 to R.D.M.
%\appendix
%\section{Floating material and so forth}

\clearpage

% Now comes the reference list.  In this document, we used \cite to call
% out citations, so we must use \bibitem in the reference list, which
% means we use the LaTeX thebibliography environment.  Please note that
% \begin{thebibliography} is followed by a null argument.  If you forget
% this, mayhem ensues, and LaTeX will say "Perhaps a missing item?" when
% you run it.  Do not call us, do not send mail when this happens.  Put
% the silly {} after the \begin{thebibliography}.
%
% Each reference has a \bibitem command to define the citation format
% to be placed in the text (in []) and the symbolic tag used for 
% cross referencing (in {}).
%
% See sample1.tex, or the AASTeX guide, for an alternative to the \cite-
% \bibitem command.

\clearpage

\begin{figure}
\plotone{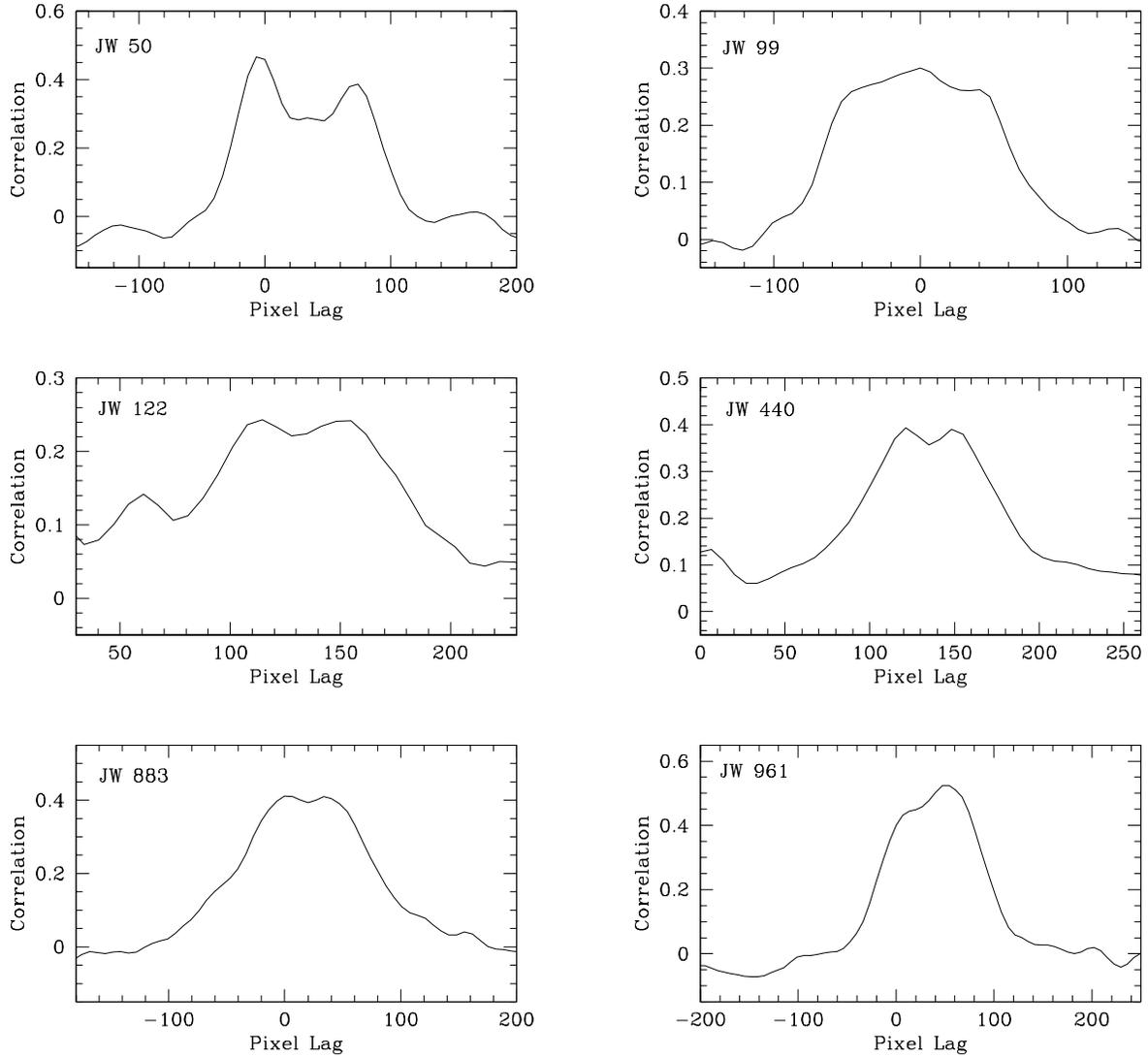}
\caption{Plotted here are the cross-correlation function peaks for six
of the seven double-lined spectroscopic binary candidates found in our
data set.  (The other SB2 candidate, JW~669, is shown in
Figure~\ref{fig:three ccfs}.)  These cross-correlation peaks show
structure (sometimes appearing as multiple overlapping Gaussians) and are
wider than typical peaks, possibly due to the presence of two sets of
spectral lines in the target star spectra.  \protect\label{fig:six
ccfs}}
\end{figure}

\begin{figure}
\plotone{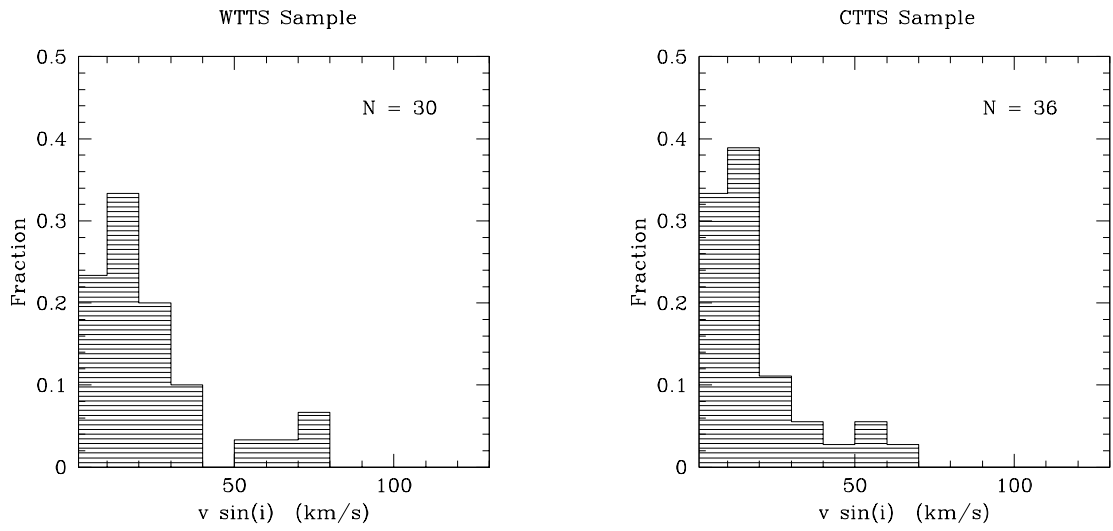}
\caption{Comparison of the \vsini\ distributions for a sample of 36
stars likely to have accretion disks (CTTS) and 30 stars not likely to
have disks (WTTS).  The fraction of slow rotators is slightly higher
for the CTTS sample, but a K-S test shows that the distributions do
not differ significantly.  \protect\label{fig:compare WTTS and CTTS}}
\end{figure}

\begin{figure}
\plotone{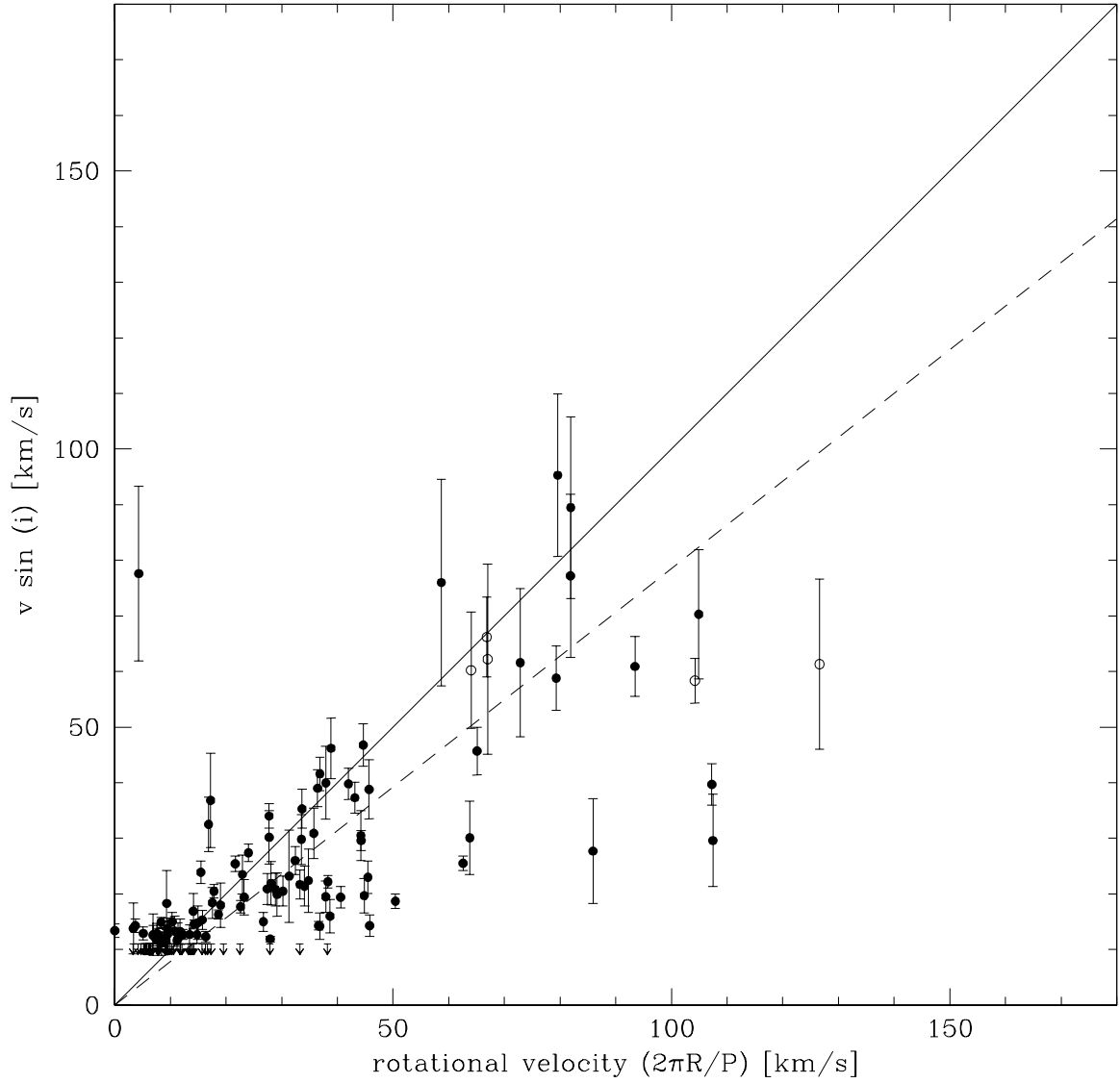}
\caption{A new version of Figure~\ref{fig:vsini vs. veq}: measured
\vsini\ versus equatorial rotation velocity is shown for 153 stars.
Here, the equatorial velocities have been calculated using
$T_{\rm{eff}}$ values that are 600~K hotter than published values.
The points scatter around the dotted \vsini\ $=$ $(\pi/4)v_{eq}$ line,
as expected for a group of stars with randomly-oriented rotation axes.
\protect\label{fig:vsini vs. veq newT}}
\end{figure}

\clearpage

\begin{deluxetable}{lccccrccccccccc}
%\tablenum{1}
%this tablewidth command makes the table set itself to the correct width
\tablewidth{0pt}
%\rotate
\tabletypesize{\scriptsize}
%footnotesize script, suggested in AASTeX help, doesn't work
\setlength{\tabcolsep}{0.02in}
\tablecaption{ONC Stars Observed for this Study}
\tablehead{\colhead{JW} & \colhead{Sam}  & \colhead{Mem} & \colhead{$I$} & \colhead{$V$$-$$I$} & \colhead{log(L)} & \colhead{log(T$_{\rm{eff}}$)} & \colhead{$\Delta$($I$$-$$K$)} & \colhead{$W_{\lambda}$(CaII)} & \colhead{Per} & \colhead{Src} & \colhead{$v$~$sin(i)$} & \colhead{$R~sin(i)$} & \colhead{SB2?} & \colhead{Date Obs}\\
\colhead{} & \colhead{} & \colhead{ (\%)} & \colhead{(mag)} & \colhead{(mag)} & \colhead{(L$_{\sun}$)} & \colhead{(K)} & \colhead{(mag)} & \colhead{(\AA)} & \colhead{(days)} & \colhead{} & \colhead{(km~s$^{-1}$)} & \colhead{(R$_{\sun}$)} & \colhead{} & \colhead{}}
\startdata
%  JW  Samp  Mem    I    V-I   logL   logT   D(I-K)    EW(CaII)  Per  Src     vsini    vserr    Rsini  SB   When_obs
   2 & C & 97 & 12.2 & 1.61 &  0.26 & 3.643 &    0.42 &  .... &  .... & ... & $\leq$11.0 & .... & .. & Jan 97,Dec 97\\
   3 & C &  0 & 13.0 & 2.50 &  0.24 & 3.580 &    0.55 &   2.0 &  3.43 &   E &  29.8 $\pm$  4.5 & 2.02 & .. &        Dec 97\\
  15 & C & 99 & 14.1 & 2.66 & $-$0.18 & 3.562 &    0.11 &   1.7 &  9.56 &   E & $\leq$11.0 & .... & .. &        Dec 97\\
  17 & C & 99 & 13.5 & 3.00 & $-$0.20 & 3.509 &    0.10 &   0.8 &  3.14 &   E &  19.9 $\pm$  3.9 & 1.23 & .. &        Dec 98\\
  18 & C & 99 & 14.4 & 2.66 & $-$0.46 & 3.544 &    0.54 &   0.6 &  .... & ... &  13.1 $\pm$  3.3 & .... & .. &        Dec 97\\
  20 & C & 99 & 13.8 & 2.67 & $-$0.40 & 3.518 &    0.10 &   0.7 &  0.67 &   E &  70.3 $\pm$ 11.6 & 0.93 & .. &        Dec 98\\
  22 & C & 99 & 13.9 & 2.38 & $-$0.34 & 3.555 &    0.14 &   2.2 &  .... & ... & $\leq$11.0 & .... & .. &        Dec 97\\
  25 & C & 99 & 15.2 & 2.54 & $-$0.88 & 3.500 &    0.52 &   0.0 &  2.28 &   E &  18.0 $\pm$  4.0 & 0.81 & .. &        Dec 98\\
  27 & C & 93 & 11.1 & 1.31 &  0.72 & 3.695 &    0.22 &  .... &  .... & ... &  38.7 $\pm$  2.2 & .... & .. &        Dec 97\\
  29 & C & 97 & 11.7 & 1.60 &  0.67 & 3.695 &    0.32 &  .... &  .... & ... &  23.8 $\pm$  1.4 & .... & .. &        Jan 97\\
  31 & C & 99 & 14.2 & 2.44 & $-$0.58 & 3.526 &    0.01 &   1.4 &  .... & ... & $\leq$11.0 & .... & .. &        Dec 98\\
  36 & C & 99 & 14.3 & 3.08 & $-$0.30 & 3.526 &    0.46 &  $-$4.1 &  2.72 &   E &  21.9 $\pm$  3.9 & 1.18 & .. &        Dec 98\\
  37 & C & 92 & 12.9 & 1.68 &  0.14 & 3.679 &    0.62 &  $-$8.4 &  8.27 &   E &  14.9 $\pm$  0.9 & 2.43 & .. &        Jan 97\\
  39 & C & 99 & 14.3 & 2.40 & $-$0.63 & 3.526 &    0.14 &   0.8 &  .... & ... &  11.7 $\pm$  1.1 & .... & .. &        Dec 98\\
  46 & C & 97 & 11.1 & 1.27 &  0.73 & 3.708 &    0.10 &  .... &  .... & ... &  14.6 $\pm$  0.6 & .... & .. &        Dec 97\\
  50 & C & 98 & 11.7 & 1.75 &  0.44 & 3.602 &    1.37 & $-$14.6 &  .... & ... &  16.4 $\pm$  2.3 & .... &  Y &        Dec 98\\
  51 & C & 99 & 14.4 & 2.37 & $-$0.57 & 3.555 &    1.05 &   0.0 &  .... & ... &  12.4 $\pm$  1.5 & .... & .. &        Dec 98\\
  55 & C & 95 & 14.4 & 2.67 & $-$0.60 & 3.526 &    0.36 &   1.0 &  .... & ... & $\leq$11.0 & .... & .. &        Dec 97\\
  63 & C & 99 & 12.8 & 2.06 &  0.26 & 3.623 &    0.20 &   1.9 &  .... & ... &  17.4 $\pm$  1.2 & .... & .. &        Jan 97\\
  65 & P & 99 & 13.8 & 2.25 & $-$0.24 & 3.580 &    0.50 &   0.0 &  7.39 &   V &  12.0 $\pm$  1.1 & 1.75 & .. &        Dec 97\\
  73 & C & 99 & 14.0 & 1.85 & $-$0.59 & 3.562 &    1.65 & $-$34.0 &  .... & ... & ........ & .... & .. &        Dec 98 \\
  75 & P & 99 & 11.1 & 1.61 &  0.89 & 3.695 &    0.09 &  .... &  3.45 &   V &  29.6 $\pm$  1.8 & 2.02 & .. &        Jan 97\\
  76 & P & 99 & 13.6 & 2.97 & $-$0.23 & 3.509 &    0.16 &   1.3 &  6.33 &   V & $\leq$11.0 & .... & .. &        Jan 97\\
  77 & P & 99 & 13.6 & 2.16 & $-$0.21 & 3.580 &    0.42 &   1.8 &  1.50 & EVS &  38.8 $\pm$  5.3 & 1.15 & .. &        Dec 97\\
  83 & P & 99 & 14.7 & 3.12 & $-$0.29 & 3.544 &    0.37 &  $-$9.2 &  7.72 &   V &  18.3 $\pm$  5.9 & 2.79 & .. &        Dec 97\\
  91 & C & 99 & 13.5 & 2.44 & $-$0.24 & 3.509 &   $-$0.06 &   1.2 & 17.08 &   E &  12.9 $\pm$  1.2 & 4.35 & .. &        Dec 97\\
  95 & C & 95 & 12.8 & 1.77 & $-$0.12 & 3.591 &    0.22 &   0.0 &  .... & ... &  61.5 $\pm$  9.0 & .... & .. &        Dec 97\\
  98 & C & 99 & 14.6 & 3.20 & $-$0.51 & 3.509 &    0.17 &   0.0 &  2.34 &   E &  20.9 $\pm$  2.8 & 0.97 & .. &        Dec 98\\
  99 & P & 99 & 13.1 & 2.51 &  0.13 & 3.562 &    0.50 &   1.4 &  1.70 &   V &  60.2 $\pm$ 10.5 & 2.02 &  Y & Jan 97,Dec 97\\
 106 & C & 99 & 13.4 & 2.42 & $-$0.29 & 3.526 &    0.10 &   1.5 &  1.70 &   E &  23.0 $\pm$  2.9 & 0.77 & .. &        Dec 98\\
 111 & C & 98 & 13.0 & 1.83 & $-$0.15 & 3.591 &    0.24 &   2.2 &  4.94 &   E & $\leq$11.0 & .... & .. &        Jan 97\\
 114 & P & 99 & 15.3 & 2.14 & $-$1.01 & 3.562 &    2.01 &  $-$1.4 &  8.73 &   V &  13.8 $\pm$  4.6 & 2.38 & .. &        Dec 97\\
 116 & P & 99 & 12.1 & 1.44 &  0.20 & 3.643 &    0.09 &   1.6 &  2.34 &   V &  39.0 $\pm$  3.3 & 1.80 & .. &        Jan 97\\
 117 & C & 99 & 13.2 & 2.13 & $-$0.05 & 3.580 &    0.77 &  $-$6.3 &  8.87 &   E & $\leq$11.0 & .... & .. &        Jan 97\\
 122 & C & 98 & 14.4 & 3.02 & $-$0.54 & 3.494 &   $-$0.17 &   0.0 &  0.98 &   E &  62.2 $\pm$ 17.1 & 1.20 &  Y &        Dec 97\\
 124 & C & 99 & 13.9 & 2.76 & $-$0.43 & 3.518 &   $-$0.02 &   0.8 &  .... & ... &  14.7 $\pm$  3.8 & .... & .. &        Dec 97\\
 126 & C & 99 & 14.4 & 2.55 & $-$0.69 & 3.526 &    0.01 &   0.8 &  2.78 &   E &  18.4 $\pm$  2.8 & 1.01 & .. &        Dec 97\\
 127 & C & 99 & 14.3 & 2.94 & $-$0.45 & 3.518 &    1.05 &  $-$2.8 &  .... & ... &  13.9 $\pm$  2.1 & .... & .. &        Dec 97\\
 128 & P & 97 & 12.6 & 2.82 &  0.28 & 3.535 &    0.32 &   0.0 &  8.83 &   V & $\leq$11.0 & .... & .. &        Dec 97\\
 133 & P & 99 & 13.2 & 2.08 & $-$0.25 & 3.555 &    0.26 &   1.6 &  2.02 &   V &  30.9 $\pm$  4.5 & 1.23 & .. &        Dec 97\\
 135 & P & 99 & 14.4 & 2.93 & $-$0.46 & 3.526 &    1.11 &  $-$2.3 &  3.69 &   V &  36.8 $\pm$  8.5 & 2.68 & .. &        Dec 97\\
 136 & P & 99 & 13.8 & 2.81 &  0.06 & 3.571 &    .... &  .... &  8.65 &   V &  11.6 $\pm$  1.7 & 1.98 & .. &        Dec 98\\
 140 & C & 99 & 13.9 & 3.10 & $-$0.19 & 3.518 &   $-$0.29 &   3.8 &  4.58 &   E & $\leq$11.0 & .... & .. &        Dec 98\\
 146 & C & 99 & 13.5 & 2.27 & $-$0.33 & 3.544 &    0.32 &   0.0 &  .... & ... &  11.2 $\pm$  1.3 & .... & .. &        Dec 98\\
 148 & P & 99 & 15.4 & 3.73 &  0.07 & 3.580 &    0.08 &   1.6 &  1.37 &  VS & ........ & .... & .. &        Dec 97 \\
 149 & P & 99 & 15.6 & 3.25 & $-$0.70 & 3.526 &    0.24 & $-$23.7 &  2.83 &   V & ........ & .... & .. &        Dec 97 \\
 150 & C & 99 & 14.0 & 2.87 &  0.27 & 3.623 &    1.13 &  $-$1.7 &  .... & ... & $\leq$11.0 & .... & .. &        Dec 98\\
 152 & C & 99 & 14.1 & 2.78 & $-$0.42 & 3.526 &    0.39 &   2.6 &  .... & ... & ........ & .... & .. &        Dec 98 \\
 157 & P & 99 & 10.2 & 1.56 &  1.19 & 3.679 &    0.11 &  .... & 17.40 &   V & $\leq$11.0 & .... & .. &        Dec 98\\
 158 & P & 99 & 13.3 & 2.15 & $-$0.23 & 3.555 &    0.71 &   1.5 &  1.95 &   V &  19.5 $\pm$  2.8 & 0.75 & .. &        Jan 97\\
 159 & P & 99 & 13.9 & 2.46 & $-$0.37 & 3.544 &   $-$1.72 &   0.3 &  1.12 &   V &  76.0 $\pm$ 18.6 & 1.68 & .. &        Dec 97\\
 163 & C & 99 & 11.8 & 1.30 &  0.22 & 3.643 &    0.53 &   1.6 &  .... & ... & $\leq$11.0 & .... & .. &        Dec 97\\
 164 & C & 99 & 14.0 & 2.38 & $-$0.47 & 3.544 &    0.63 &  $-$0.7 &  6.52 &   E &  12.2 $\pm$  1.6 & 1.57 & .. &        Dec 98\\
 165 & P & 97 & 11.7 & 1.82 &  1.49 & 3.897 &    0.11 &  .... &  5.77 &   V &  17.7 $\pm$  1.1 & 2.02 & .. &        Dec 98\\
 166 & C & 99 & 14.2 & 2.81 & $-$0.46 & 3.494 &   $-$0.26 &   0.0 &  0.67 &   E &  29.6 $\pm$  8.3 & 0.39 & .. &        Dec 97\\
 167 & P & 99 & 14.9 & 3.08 & $-$0.03 & 3.602 &    0.86 &   1.4 &  3.43 &   E & $\leq$11.0 & .... & .. &        Jan 97\\
 169 & P & 99 & 15.9 & .... & $-$1.30 & 3.526 &    .... &  .... &  3.85 &   V & ........ & .... & .. &        Dec 98 \\
 174 & P & 99 & 14.1 & 2.40 & $-$0.59 & 3.535 &    0.12 &   3.7 &  1.36 & EVS &  46.2 $\pm$  5.4 & 1.24 & .. & Jan 97,Dec 97\\
 175 & P & 99 & 14.1 & 2.41 & $-$0.55 & 3.526 &    0.56 &   0.0 &  9.19 &   V & $\leq$11.0 & .... & .. &        Dec 97\\
 176 & C & 99 & 11.6 & 1.89 &  0.47 & 3.591 &    0.54 &   0.8 &  .... & ... & $\leq$11.0 & .... & .. &        Dec 98\\
 180 & C & 99 & 14.4 & 2.94 & $-$0.54 & 3.494 &    0.05 &   0.2 &  .... & ... &  41.9 $\pm$  9.6 & .... & .. &        Dec 98\\
 187 & C & 99 & 14.3 & 2.85 &  0.22 & 3.643 &    0.53 &   1.4 & 14.41 &   E & $\leq$11.0 & .... & .. &        Dec 98\\
 191 & P & 99 & 14.2 & 2.78 & $-$0.37 & 3.535 &    0.18 &   0.0 &  8.63 &   V & $\leq$11.0 & .... & .. &        Dec 97\\
 203 & C & 98 & 13.8 & 2.48 & $-$0.11 & 3.580 &    0.63 &   1.0 &  .... & ... & $\leq$11.0 & .... & .. &        Dec 97\\
 208 & C & 96 & 14.3 & 2.52 & $-$0.56 & 3.535 &    0.24 &   0.0 &  .... & ... & $\leq$11.0 & .... & .. &        Dec 98\\
 211 & C & 99 & 14.0 & 2.61 & $-$0.42 & 3.535 &    0.24 &   1.4 &  5.46 &   E &  13.2 $\pm$  2.3 & 1.42 & .. &        Dec 97\\
 220 & P & 99 & 12.7 & 2.25 &  0.07 & 3.555 &    0.67 &   1.0 &  2.33 &   V &  19.7 $\pm$  3.1 & 0.91 & .. &        Jan 97\\
 221 & C & 98 & 11.3 & 2.66 &  1.41 & 3.679 &   $-$0.01 &   1.5 &  .... & ... &  56.9 $\pm$  5.3 & .... & .. &        Jan 97\\
 222 & P & 98 & 14.1 & 2.85 & $-$0.23 & 3.544 &    0.71 &   1.2 &  5.17 &   V &  14.8 $\pm$  3.0 & 1.51 & .. &        Dec 97\\
 225 & C & 99 & 13.4 & 2.30 & $-$0.18 & 3.555 &    0.14 &   0.5 &  .... & ... &  25.8 $\pm$  3.5 & .... & .. &        Jan 97\\
 229 & C & 99 & 14.2 & 2.41 & $-$0.43 & 3.555 &    0.37 &   2.1 &  7.75 &   E &  12.9 $\pm$  1.8 & 1.98 & .. &        Dec 97\\
 232 & P & 79 & 11.7 & 1.86 &  0.82 & 3.695 &    0.23 &  .... &  5.08 &   V &  34.0 $\pm$  2.2 & 3.41 & .. &        Dec 97\\
 237 & C & 99 & 15.3 & 3.22 & $-$0.78 & 3.509 &    0.75 &   0.0 &  5.22 &   E & $\leq$11.0 & .... & .. &        Dec 98\\
 239 & P & 99 & 12.7 & 2.25 &  0.05 & 3.555 &    0.38 &   1.5 &  4.45 &   V &  23.5 $\pm$  3.5 & 2.07 & .. &        Jan 97\\
 243 & P & 99 & 13.6 & 2.36 & $-$0.31 & 3.544 &    0.64 &   1.2 & 10.20 &   V &  12.5 $\pm$  1.8 & 2.52 & .. &        Dec 97\\
 248 & P & 99 & 12.7 & 2.17 &  0.10 & 3.571 &    0.59 &  $-$0.9 &  6.86 &   V &  12.7 $\pm$  1.6 & 1.72 & .. &        Jan 97\\
 249 & C & 99 & 13.6 & 2.35 &  0.35 & 3.695 &    1.14 &   1.7 &  .... & ... &  25.9 $\pm$  1.7 & .... & .. &        Dec 97\\
 250 & P & 99 & 14.0 & 3.00 & $-$0.24 & 3.526 &    0.27 &   0.9 &  2.71 &   V &  20.5 $\pm$  2.7 & 1.10 & .. &        Dec 97\\
 254 & P & 99 & 14.3 & 3.21 & $-$0.24 & 3.526 &    0.50 &   5.6 &  3.52 &   V &  19.4 $\pm$  3.2 & 1.35 & .. &        Dec 98\\
 256 & C & 99 & 14.1 & 1.13 & $-$0.52 & 3.518 &    .... &   2.1 &  .... & ... & $\leq$11.0 & .... & .. &        Dec 98\\
 272 & C & 99 & 14.1 & 2.30 & $-$0.46 & 3.555 &    0.72 &  $-$2.2 &  .... & ... & $\leq$11.0 & .... & .. &        Dec 98\\
 275 & P & 99 & 13.0 & 2.28 & $-$0.13 & 3.544 &    0.76 &   1.8 & 20.10 &   V &  77.6 $\pm$ 15.7 & .... & .. &        Dec 98\\
 284 & P & 99 & 14.6 & 3.11 & $-$0.41 & 3.526 &    1.18 &  $-$2.0 &  3.11 &  EV & ........ & .... & .. &        Jan 97 \\
 291 & C & 99 & 12.9 & 2.07 & $-$0.06 & 3.562 &    0.32 &   1.5 &  1.91 &   E &  14.3 $\pm$  1.9 & 0.54 & .. &        Dec 98\\
 305 & C & 99 & 14.6 & 2.39 & $-$0.72 & 3.518 &    0.42 &   0.0 &  .... & ... &  13.8 $\pm$  2.5 & .... & .. &        Dec 98\\
 308 & P &  0 & 12.6 & 2.54 &  0.06 & 3.526 &    .... &   0.7 &  4.17 &   V &  30.2 $\pm$  4.8 & 2.49 & .. &        Dec 98\\
 310 & C & 98 & 13.8 & 3.57 & $-$0.05 & 3.500 &    0.58 &   0.8 &  .... & ... &  54.5 $\pm$ 15.2 & .... & .. &        Dec 98\\
 311 & P & 99 & 14.9 & 3.13 & $-$0.60 & 3.518 &    0.22 &   0.0 &  6.14 &   V & $\leq$11.0 & .... & .. &        Dec 98\\
 315 & P & 99 & 14.8 & 2.85 & $-$0.73 & 3.518 &   $-$0.03 &   0.0 &  8.85 &   V & $\leq$11.0 & .... & .. &        Dec 97\\
 320 & P & 99 & 14.7 & 2.37 & $-$0.08 & 3.695 &    0.98 &   0.3 &  5.41 &   V &  11.6 $\pm$  2.0 & 1.24 & .. &        Dec 97\\
 321 & C & 95 & 13.5 & 2.09 & $-$0.19 & 3.580 &    .... &  $-$5.1 &  7.44 &   E &  13.8 $\pm$  1.3 & 2.03 & .. &        Dec 97\\
 328 & C & 99 & 12.9 & 1.64 &  0.09 & 3.661 &    1.02 &   0.0 & 18.83 &   E &  14.3 $\pm$  1.2 & 5.32 & .. &        Dec 97\\
 330 & P & 99 & 11.8 & 1.42 &  0.45 & 3.679 &    0.08 &   1.8 &  1.57 &   V &  25.5 $\pm$  1.3 & 0.79 & .. &        Jan 97\\
 332 & C & 99 & 14.2 & 2.07 & $-$0.59 & 3.518 &    2.46 &  $-$0.9 &  .... & ... & $\leq$11.0 & .... & .. &        Dec 98\\
 333 & C & 99 & 14.7 & 1.92 & $-$0.56 & 3.483 &    0.54 &   3.7 &  .... & ... & $\leq$11.0 & .... & .. &        Dec 98\\
 334 & P & 99 & 13.6 & 2.35 & $-$0.20 & 3.562 &    1.20 &  $-$4.8 &  5.27 &   V &  16.9 $\pm$  3.2 & 1.76 & .. &        Dec 97\\
 335 & C & 99 & 15.6 & 1.31 & $-$1.21 & 3.562 &    3.10 & $-$26.0 &  .... & ... & ........ & .... & .. &        Dec 98 \\
 337 & P & 99 & 11.7 & 1.76 &  0.34 & 3.591 &    0.52 &   1.9 & 19.50 &   V & $\leq$11.0 & .... & .. &        Dec 97\\
 345 & P & 81 & 12.4 & 2.38 &  0.37 & 3.571 &    0.42 &   0.0 &  8.21 &   V &  32.5 $\pm$  4.9 & 5.27 & .. &        Dec 97\\
 347 & P & 99 & 12.5 & 1.75 &  0.09 & 3.602 &    0.81 &   0.0 &  7.33 &  VS & $\leq$11.0 & .... & .. &        Dec 97\\
 348 & C & 99 & 11.2 & 1.53 &  0.92 & 3.719 &    1.00 &  .... &  .... & ... &  21.7 $\pm$  1.1 & .... & .. &        Dec 98\\
 352 & P & 99 & 10.4 & 2.10 &  1.09 & 3.591 &    .... &   1.9 &  8.00 &   V &  14.3 $\pm$  0.8 & 2.26 & .. &        Jan 97\\
 362 & P & 99 & 13.9 & 3.23 & $-$0.29 & 3.500 &    0.38 &   0.0 &  2.73 &   V &  23.2 $\pm$  8.3 & 1.25 & .. &        Dec 97\\
 363 & C & 98 & 12.7 & 2.51 &  0.12 & 3.544 &    0.40 &   1.1 &  .... & ... &  33.4 $\pm$  7.6 & .... & .. &        Jan 97\\
 365 & P & 99 & 11.7 & 1.96 &  0.82 & 3.679 &    0.40 &  .... &  4.08 &   V &  41.6 $\pm$  3.0 & 3.35 & .. &        Jan 97\\
 373 & C & 99 & 13.5 & 3.37 &  0.99 & 3.679 &    0.20 &   2.1 &  9.81 &   E &  16.3 $\pm$  0.8 & 3.16 & .. &        Dec 98\\
 378 & P & 99 & 12.9 & 2.11 &  0.05 & 3.580 &    1.08 &   1.5 &  9.03 &   V & $\leq$11.0 & .... & .. &        Dec 97\\
 379 & P & 99 & 15.2 & 3.87 & $-$0.58 & 3.483 &    0.07 &  .... & 11.30 &  VS & $\leq$11.0 & .... & .. &        Dec 97\\
 381 & P & 99 & 13.2 & 2.11 & $-$0.14 & 3.571 &    1.26 &   0.0 & 16.20 &   V & $\leq$11.0 & .... & .. &        Dec 97\\
 388 & P & 99 & 13.6 & 2.62 &  0.22 & 3.602 &    0.06 &   1.2 &  9.08 &   V &  11.9 $\pm$  0.9 & 2.13 & .. &        Dec 97\\
 398 & C & 99 & 14.5 & 2.44 & $-$0.41 & 3.580 &    0.30 &   2.2 &  .... & ... & $\leq$11.0 & .... & .. &        Dec 98\\
 406 & P & 99 & 13.9 & 3.41 &  0.37 & 3.562 &    0.31 &   1.7 &  2.25 &   V &  30.1 $\pm$  6.6 & 1.34 & .. &        Dec 97\\
 417 & P & 98 & 13.9 & 2.44 & $-$0.22 & 3.571 &    0.56 &   0.0 &  7.41 &   V & $\leq$11.0 & .... & .. &        Dec 97\\
 421 & C & 99 & 11.5 & 1.46 &  0.47 & 3.643 &    0.80 &  $-$3.5 &  .... & ... & $\leq$11.0 & .... & .. &        Dec 98\\
 429 & C & 99 & 12.7 & 1.75 &  0.57 & 3.753 &    .... &   4.5 &  .... & ... &  13.2 $\pm$  0.7 & .... & .. &        Dec 98\\
 433 & C & 93 & 11.9 & 1.59 &  0.43 & 3.661 &    0.21 &  .... &  .... & ... & $\leq$11.0 & .... & .. & Jan 97,Dec 97\\
 437 & P & 99 & 11.8 & 2.07 &  0.56 & 3.602 &    0.53 &   1.4 &  2.34 &   V &  45.7 $\pm$  4.3 & 2.11 & .. &        Jan 97\\
 439 & P & 99 & 15.0 & 3.33 & $-$0.12 & 3.562 &    0.62 &   0.7 &  8.30 &   V & $\leq$11.0 & .... & .. &        Dec 98\\
 440 & C & 99 & 14.5 & 3.09 & $-$0.44 & 3.518 &   $-$0.35 &   0.0 &  .... & ... &  49.3 $\pm$  8.3 & .... &  Y &        Dec 98\\
 443 & C & 99 & 14.0 & 0.95 & $-$0.55 & 3.544 &    0.84 &  $-$1.5 &  .... & ... &  11.5 $\pm$  2.2 & .... & .. &        Dec 98\\
 454 & P & 99 & 11.9 & 2.10 &  0.78 & 3.661 &    0.39 &  $-$1.0 &  8.67 &   V &  20.5 $\pm$  1.2 & 3.51 & .. &        Dec 97\\
 460 & C & 95 & 11.7 & 1.58 &  0.65 & 3.695 &    0.88 &  .... &  .... & ... &  40.9 $\pm$  4.2 & .... & .. &        Dec 98\\
 466 & P & 99 & 14.4 & 2.88 & $-$0.56 & 3.518 &    0.35 &  .... &  6.17 &   V &  12.7 $\pm$  3.0 & 1.55 & .. &        Dec 97\\
 470 & P & 99 & 12.5 & 2.63 &  0.84 & 3.661 &    .... &   6.1 & 10.70 &   V &  23.9 $\pm$  2.0 & 5.05 & .. &        Dec 98\\
 478 & P & 99 & 11.6 & 2.28 &  0.69 & 3.580 &    0.43 &  .... &  5.74 &   V &  35.3 $\pm$  3.5 & 4.00 & .. &        Dec 97\\
 479 & P & 99 & 11.0 & 1.80 &  1.16 & 3.719 &    0.57 &   2.0 &  8.71 &   V &  25.4 $\pm$  1.4 & 4.37 & .. &        Dec 98\\
 481 & P & 99 & 14.7 & 2.99 & $-$0.52 & 3.526 &    0.35 &   0.0 &  6.56 &   V & ........ & .... & .. &        Dec 97 \\
 482 & C & 99 & 13.3 & 2.38 &  0.10 & 3.591 &    0.60 & $-$50.3 &  .... & ... &  13.9 $\pm$  3.0 & .... & .. &        Dec 98\\
 485 & P & 99 & 14.3 & 2.49 & $-$0.42 & 3.555 &    1.18 &   0.8 &  9.66 &   V & $\leq$11.0 & .... & .. &        Dec 97\\
 487 & C & 99 & 12.8 & 1.78 &  0.07 & 3.623 &    0.28 &   1.8 &  .... & ... &  12.5 $\pm$  0.9 & .... & .. &        Jan 97\\
 498 & P & 99 & 13.8 & 2.08 & $-$0.44 & 3.562 &    .... &  .... &  7.42 &   V &  11.7 $\pm$  1.4 & 1.72 & .. &        Dec 97\\
 501 & C & 99 & 13.2 & 1.82 & $-$0.28 & 3.580 &    0.62 &   1.1 &  .... & ... & $\leq$11.0 & .... & .. &        Dec 98\\
 502 & C & 99 & 13.5 & 2.57 & $-$0.22 & 3.535 &    0.97 &   0.0 &  .... & ... &  23.8 $\pm$  4.7 & .... & .. &        Dec 98\\
 526 & P & 98 & 11.9 & 1.90 &  0.52 & 3.623 &    0.74 &   1.2 &  1.68 &   V &  95.3 $\pm$ 14.6 & 3.16 & .. &        Jan 97\\
 527 & C & 99 & 13.0 & 2.09 & $-$0.06 & 3.571 &    0.63 &   0.8 &  .... & ... &  12.7 $\pm$  1.5 & .... & .. &        Dec 98\\
 534 & C & 85 & 12.7 & 1.43 & $-$0.03 & 3.544 &    1.33 & $-$69.5 &  .... & ... & ........ & .... & .. &        Dec 98 \\
 544 & P & 99 & 10.8 & 2.22 &  1.30 & 3.670 &   $-$0.17 &   1.6 &  3.42 &   V &  58.8 $\pm$  5.8 & 3.97 & .. &        Jan 97\\
 565 & C & 99 & 14.5 & 2.95 & $-$0.62 & 3.509 &    0.22 &   0.0 &  .... & ... & $\leq$11.0 & .... & .. &        Dec 98\\
 567 & P & 99 &  9.7 & 1.73 &  1.43 & 3.661 &   $-$0.17 &  $-$1.7 &  8.53 &   V &  22.2 $\pm$  1.1 & 3.74 & .. &        Dec 97\\
 576 & P & 99 & 13.1 & 2.28 & $-$0.07 & 3.555 &    0.23 &   0.0 &  2.01 &   V &  30.5 $\pm$  4.5 & 1.21 & .. &        Dec 97\\
 585 & C & 99 & 11.5 & 1.37 &  0.39 & 3.602 &    .... &   1.8 &  .... & ... &  18.4 $\pm$  0.9 & .... & .. &        Dec 98\\
 589 & P & 99 & 11.3 & 2.12 &  1.23 & 3.719 &    0.11 &  $-$0.6 & 14.30 &   V &  14.6 $\pm$  0.7 & 4.12 & .. &        Dec 98\\
 592 & C & 99 & 14.4 & 2.58 & $-$0.57 & 3.535 &    0.06 &   0.0 &  7.92 &   E & $\leq$11.0 & .... & .. &        Dec 98\\
 601 & P & 98 & 12.5 & 1.25 &  0.05 & 3.679 &    0.05 &   2.5 &  2.22 &   V &  11.9 $\pm$  0.5 & 0.52 & .. &        Jan 97\\
 625 & C & 99 & 14.4 & 1.07 & $-$0.73 & 3.562 &    1.37 &   6.1 &  .... & ... & ........ & .... & .. &        Dec 98 \\
 629 & C & 99 & 13.5 & 4.34 &  1.47 & 3.643 &    1.41 & $-$16.9 &  .... & ... &  56.3 $\pm$ 11.3 & .... & .. &        Dec 98\\
 631 & C & 99 & 12.9 & 2.20 &  0.22 & 3.602 &    1.29 &   1.6 &  0.96 &   E &  39.7 $\pm$  3.7 & 0.75 & .. &        Dec 98\\
 636 & P & 99 & 13.4 & 2.46 &  0.04 & 3.575 &    0.79 &  .... &  5.59 &   V & $\leq$11.0 & .... & .. &        Jan 97\\
 639 & P & 99 & 13.6 & 2.53 & $-$0.37 & 3.526 &    .... &  .... &  5.24 &   V &  12.7 $\pm$  1.7 & 1.31 & .. &        Dec 97\\
 641 & P & 99 & 11.5 & 1.74 &  0.91 & 3.719 &    0.12 &   0.0 &  3.17 &   V &  46.8 $\pm$  3.8 & 2.93 & .. &        Jan 97\\
 647 & C & 99 & 13.7 & 2.31 & $-$0.22 & 3.494 &    .... &  $-$5.0 &  .... & ... &  13.8 $\pm$  1.3 & .... & .. &        Dec 98\\
 648 & P & 99 & 13.2 & 2.58 &  0.24 & 3.580 &    .... &   0.9 & 34.50 &   V & $\leq$11.0 & .... & .. &        Dec 98\\
 661 & C & 99 & 13.5 & 1.57 & $-$0.39 & 3.580 &    1.80 &   2.1 &  .... & ... & $\leq$11.0 & .... & .. &        Dec 97\\
 662 & C & 95 & 14.4 & 1.78 & $-$0.57 & 3.623 &    1.36 &  $-$1.0 &  .... & ... &  30.7 $\pm$  4.8 & .... & .. &        Dec 98\\
 664 & P & 99 & 14.1 & 2.88 &  0.26 & 3.623 &   $-$0.18 &   1.4 &  7.20 &   V & $\leq$11.0 & .... & .. &        Dec 97\\
 669 & P & 99 & 10.8 & 1.67 &  1.12 & 3.708 &    0.14 &   1.8 &  1.81 &   V &  58.4 $\pm$  4.0 & 2.09 &  Y & Jan 97,Dec 97\\
 673 & P & 99 & 13.1 & 1.83 &  0.06 & 3.643 &    0.81 &   1.3 &  1.44 &  EV &  18.7 $\pm$  1.3 & 0.53 & .. &        Jan 97\\
 678 & P & 99 & 12.3 & 1.84 &  0.55 & 3.695 &    0.24 &  .... & 13.00 &   V &  12.4 $\pm$  0.7 & 3.18 & .. &        Dec 97\\
 683 & P & 99 & 11.8 & 1.85 &  0.53 & 3.623 &    0.28 &   1.9 & 11.50 &   V & $\leq$11.0 & .... & .. &        Dec 98\\
 694 & C & 99 & 14.7 & 2.99 &  0.05 & 3.623 &    0.18 &  $-$3.8 &  .... & ... & $\leq$11.0 & .... & .. &        Dec 98\\
 701 & C & 99 & 12.4 & 1.26 &  0.25 & 3.722 &    1.27 &  $-$8.3 &  .... & ... &  12.4 $\pm$  0.8 & .... & .. &        Dec 98\\
 704 & P & 99 & 14.7 & 2.41 & $-$0.59 & 3.562 &    .... &  .... &  6.87 &   V &  12.7 $\pm$  3.7 & 1.72 & .. &        Dec 97\\
 706 & C & 99 & 11.8 & 1.34 &  0.39 & 3.679 &   $-$0.03 &  .... &  .... & ... &  12.0 $\pm$  0.6 & .... & .. &        Dec 98\\
 710 & P & 99 & 12.7 & 1.17 & $-$0.04 & 3.562 &    0.68 &  $-$2.1 &  7.59 &   V & $\leq$11.0 & .... & .. &        Dec 98\\
 716 & P & 99 & 14.4 & 2.88 & $-$0.57 & 3.500 &    1.15 &  $-$2.7 &  3.95 &   V & $\leq$11.0 & .... & .. &        Dec 98\\
 717 & P & 99 & 16.1 & .... & $-$1.39 & 3.535 &    .... &   0.0 &  4.00 &   V & ........ & .... & .. &        Dec 98 \\
 719 & C & 99 & 14.5 & 3.06 & $-$0.56 & 3.494 &    0.10 &   0.0 &  6.88 &   E & $\leq$11.0 & .... & .. &        Dec 97\\
 721 & P & 99 & 13.1 & 2.08 & $-$0.06 & 3.580 &    0.82 &  $-$3.8 &  2.45 &   V & $\leq$11.0 & .... & .. &        Dec 97\\
 727 & P & 99 & 13.9 & 2.44 & $-$0.38 & 3.544 &    0.75 &   0.8 &  6.03 &   V &  13.3 $\pm$  2.7 & 1.58 & .. &        Dec 97\\
 728 & C & 99 & 13.6 & 2.32 & $-$0.20 & 3.562 &    0.54 &   0.9 & 13.87 &   E & $\leq$11.0 & .... & .. &        Dec 97\\
 731 & P & 88 & 12.3 & 1.74 &  0.10 & 3.591 &    0.28 &   2.2 &  7.64 &   V &  12.7 $\pm$  0.9 & 1.92 & .. &        Dec 97\\
 733 & C & 99 & 13.1 & 2.33 &  0.04 & 3.571 &    0.58 &   1.7 &  3.28 &   E &  20.8 $\pm$  3.0 & 1.35 & .. &        Dec 98\\
 736 & C & 99 & 13.9 & 2.27 & $-$0.25 & 3.580 &    0.65 &   1.6 & 11.68 &   E & $\leq$11.0 & .... & .. &        Dec 98\\
 744 & C & 99 & 13.4 & 1.75 & $-$0.36 & 3.562 &    2.42 &  $-$0.9 &  .... & ... & $\leq$11.0 & .... & .. &        Dec 98\\
 751 & C & 99 & 13.5 & 2.47 & $-$0.07 & 3.562 &    0.78 &  $-$3.0 &  .... & ... & $\leq$11.0 & .... & .. &        Dec 97\\
 756 & C & 99 & 12.8 & 2.15 &  0.09 & 3.580 &    1.67 &  $-$2.0 &  .... & ... &  33.2 $\pm$  2.7 & .... & .. &        Dec 98\\
 758 & P & 99 & 14.1 & 3.45 & $-$0.15 & 3.509 &    0.48 &  $-$2.0 &  2.51 &   V &  16.0 $\pm$  3.0 & 0.79 & .. &        Dec 98\\
 765 & P & 97 & 13.2 & 2.67 & $-$0.18 & 3.522 &    0.09 &   1.3 &  2.42 &   V &  14.2 $\pm$  2.4 & 0.68 & .. &        Dec 97\\
 779 & C & 95 & 12.5 & 1.34 &  0.05 & 3.661 &    0.55 &  .... &  .... & ... &  18.7 $\pm$  1.1 & .... & .. &        Dec 98\\
 782 & C & 99 & 14.2 & 2.14 & $-$0.41 & 3.591 &    2.18 &  $-$2.5 &  .... & ... & $\leq$11.0 & .... & .. &        Dec 98\\
 786 & P & 89 & 13.9 & 2.39 & $-$0.23 & 3.571 &    1.30 &   0.0 &  8.81 &   V & $\leq$11.0 & .... & .. &        Dec 97\\
 787 & P & 99 & 12.7 & .... & $-$0.07 & 3.571 &    .... &   0.6 &  9.09 &   E & $\leq$11.0 & .... & .. &        Dec 98\\
 788 & P & 99 & 14.5 & 3.48 & $-$0.28 & 3.509 &    0.28 &   0.0 &  4.82 &  EV & $\leq$11.0 & .... & .. &        Dec 97\\
 790 & P & 79 & 10.9 & 1.06 &  0.73 & 3.719 &    0.12 &   2.2 &  2.74 &   V &  39.8 $\pm$  2.8 & 2.15 & .. &        Jan 97\\
 792 & P & 99 & 13.6 & 2.64 &  0.08 & 3.580 &    1.05 &   1.4 & 10.30 &   V & $\leq$11.0 & .... & .. &        Dec 97\\
 794 & P &  0 & 10.9 & 0.81 &  0.62 & 3.722 &    0.03 &  .... &  2.62 &   V & $\leq$11.0 & .... & .. &        Dec 98\\
 795 & P & 99 & 10.6 & 1.19 &  1.02 & 3.744 &    0.12 &   1.7 &  1.55 &   V &  60.9 $\pm$  5.4 & 1.86 & .. &        Jan 97\\
 811 & P & 99 & 13.4 & 2.56 &  0.13 & 3.580 &    1.25 &   0.0 & 11.10 &   V & $\leq$11.0 & .... & .. &        Dec 98\\
 813 & P & 99 & 12.4 & 1.75 &  0.05 & 3.580 &    0.56 &   1.9 &  2.85 &  EV &  26.0 $\pm$  2.5 & 1.46 & .. &        Dec 97\\
 815 & P & 99 & 13.8 & 2.69 & $-$0.34 & 3.526 &    0.52 &   0.0 &  6.42 &   V &  12.9 $\pm$  1.9 & 1.64 & .. &        Dec 97\\
 817 & P & 99 & 13.8 & 3.04 & $-$0.30 & 3.500 &    0.22 &  .... &  2.23 &  EV &  40.0 $\pm$  6.5 & 1.76 & .. & Jan 97,Dec 97\\
 818 & C & 98 & 12.1 & 1.82 &  0.20 & 3.591 &    0.26 &  .... &  .... & ... &  14.5 $\pm$  0.9 & .... & .. &        Dec 98\\
 819 & P & 99 & 13.2 & 2.85 & $-$0.06 & 3.518 &    0.12 &   1.3 &  1.21 &   E &  27.7 $\pm$  9.4 & 0.66 & .. &        Dec 97\\
 823 & P & 99 & 13.9 & 3.28 & $-$0.31 & 3.494 &    0.25 &   0.0 &  9.00 &   V & $\leq$11.0 & .... & .. &        Dec 98\\
 826 & P & 91 & 14.0 & 1.38 & $-$0.60 & 3.602 &    3.01 &  $-$6.3 &  9.57 &   V & $\leq$11.0 & .... & .. &        Dec 97\\
 830 & P & 99 & 13.0 & 2.36 &  0.01 & 3.555 &    0.63 &   1.8 &  9.23 &   V & $\leq$11.0 & .... & .. &        Dec 98\\
 834 & C & 99 & 13.0 & 2.24 &  .... &  .... &    .... &  .... &  7.01 &   E &  13.4 $\pm$  1.2 & 2.44 & .. &        Dec 98\\
 835 & P & 99 & 14.1 & 2.52 & $-$0.54 & 3.526 &    0.24 &   0.0 &  8.85 &   V & $\leq$11.0 & .... & .. &        Dec 97\\
 836 & P & 99 & 13.4 & 2.20 & $-$0.11 & 3.580 &    1.10 &   0.9 & 12.20 &   V & $\leq$11.0 & .... & .. &        Dec 97\\
 837 & P & 99 & 13.3 & 2.21 & $-$0.06 & 3.580 &    0.62 &   1.2 & 10.60 &   V & $\leq$11.0 & .... & .. &        Dec 97\\
 839 & P & 99 & 13.5 & 3.70 &  0.70 & 3.562 &   $-$1.41 &   1.2 &  7.52 &   V & $\leq$11.0 & .... & .. &        Dec 97\\
 843 & P & 98 & 13.8 & 2.45 & $-$0.33 & 3.544 &    0.02 &   0.0 &  0.84 &   E &  89.5 $\pm$ 16.3 & 1.48 & .. & Jan 97,Dec 97\\
 850 & P & 99 & 13.4 & 2.14 & $-$0.11 & 3.591 &    0.73 &  $-$4.1 &  7.78 &   V & $\leq$11.0 & .... & .. &        Jan 97\\
 855 & P & 99 & 14.1 & 2.34 & $-$0.51 & 3.544 &    0.79 &   0.0 &  7.08 &   V &  11.8 $\pm$  2.0 & 1.65 & .. &        Dec 97\\
 862 & C & 99 & 15.5 & 3.80 & $-$0.77 & 3.471 &    0.24 &  $-$4.5 &  3.49 &   E & ........ & .... & .. &        Dec 97 \\
 863 & P & 99 & 14.0 & 2.62 & $-$0.50 & 3.518 &    0.71 &   0.0 &  7.91 &   V & $\leq$11.0 & .... & .. &        Dec 97\\
 864 & C & 99 & 14.4 & 2.46 & $-$0.70 & 3.526 &    0.59 &   0.7 &  3.70 &   E & ........ & .... & .. &        Dec 98 \\
 866 & P & 98 & 11.8 & 2.13 &  0.99 & 3.708 &    0.18 &  .... &  6.76 &   V &  27.4 $\pm$  1.6 & 3.66 & .. &        Jan 97\\
 868 & C & 99 & 13.2 & 1.96 &  0.16 & 3.661 &    0.42 &   1.4 &  .... & ... &  13.7 $\pm$  0.8 & .... & .. &        Dec 98\\
 872 & P & 99 & 15.2 & 3.24 & $-$0.83 & 3.494 &    0.60 &   0.4 &  4.19 &   V & ........ & .... & .. &        Dec 97 \\
 880 & P & 99 & 14.2 & 2.74 & $-$0.51 & 3.509 &   $-$0.10 &   0.0 &  1.58 &   E &  19.4 $\pm$  1.9 & 0.61 & .. &        Dec 98\\
 883 & C & 99 & 13.3 & 2.62 &  0.12 & 3.562 &    .... &  .... &  0.85 &   E &  61.3 $\pm$ 15.3 & 1.03 &  Y &        Jan 97\\
 887 & C & 97 & 10.3 & 1.42 &  1.39 & 3.771 &    0.05 &  .... &  .... & ... &  66.2 $\pm$  6.3 & .... & .. &        Dec 98\\
 890 & P & 99 & 13.0 & 2.11 & $-$0.16 & 3.555 &    0.46 &   1.6 &  1.10 &   V &  61.6 $\pm$ 13.3 & 1.34 & .. &        Jan 97\\
 892 & P & 99 & 14.0 & 2.22 & $-$0.13 & 3.623 &    .... &   1.0 &  8.56 &   V &  12.0 $\pm$  0.9 & 2.03 & .. &        Dec 97\\
 906 & C & 99 & 14.2 & 2.45 & $-$0.58 & 3.526 &    0.36 &   1.6 & 10.66 &   E & $\leq$11.0 & .... & .. &        Dec 97\\
 907 & P & 99 & 12.8 & 1.60 &  0.17 & 3.679 &   $-$0.20 &   1.8 &  1.65 &   V &  37.3 $\pm$  2.8 & 1.22 & .. &        Jan 97\\
 911 & C & 99 & 12.7 & 1.62 &  0.15 & 3.661 &    1.01 &  .... &  .... & ... & $\leq$11.0 & .... & .. &        Dec 98\\
 912 & C & 99 & 13.9 & 2.54 & $-$0.48 & 3.526 &    0.48 &   1.2 &  .... & ... & $\leq$11.0 & .... & .. &        Dec 98\\
 914 & P & 99 & 13.2 & 2.09 & $-$0.23 & 3.555 &    0.71 &   0.0 &  7.71 &   V & $\leq$11.0 & .... & .. &        Dec 97\\
 919 & C & 97 & 14.4 & 2.93 & $-$0.45 & 3.526 &   $-$0.22 &   1.4 &  .... & ... & $\leq$11.0 & .... & .. &        Dec 98\\
 926 & C & 96 & 12.8 & 1.93 &  0.11 & 3.602 &    0.09 &   1.8 &  5.54 &   E &  12.3 $\pm$  1.0 & 1.35 & .. &        Dec 98\\
 927 & C & 99 & 14.0 & 2.29 & $-$0.43 & 3.509 &    0.01 &   1.0 &  2.02 &   E &  22.4 $\pm$  5.7 & 0.89 & .. &        Dec 97\\
 930 & P & 99 & 14.1 & 2.56 &  0.04 & 3.623 &   $-$0.71 &   5.2 &  2.88 &   V &  15.0 $\pm$  1.7 & 0.85 & .. &        Dec 97\\
 933 & P & 98 & 13.7 & 2.48 & $-$0.31 & 3.509 &    0.05 &   1.0 &  5.98 &   V & $\leq$11.0 & .... & .. &        Jan 97\\
 938 & C & 99 & 12.6 & 2.70 &  0.99 & 3.695 &    0.19 &   1.3 &  5.15 &   E &  21.7 $\pm$  2.6 & 2.21 & .. &        Dec 97\\
 940 & C & 95 & 12.0 & 1.63 &  0.25 & 3.535 &   $-$0.13 &  .... &  .... & ... &  61.6 $\pm$ 14.9 & .... & .. &        Dec 98\\
 945 & C & 99 & 12.5 & 2.40 &  2.29 & 4.111 &    0.58 &  .... &  .... & ... & ........ & .... & .. &        Dec 98 \\
 949 & C & 90 & 12.3 & 2.55 &  0.47 & 3.562 &    0.55 &   0.0 &  .... & ... &  67.3 $\pm$ 18.8 & .... & .. &        Dec 97\\
 961 & C & 98 & 12.3 & 1.71 &  0.19 & 3.612 &    0.18 &   1.9 &  1.43 &   E &  66.2 $\pm$  7.2 & 1.87 &  Y &        Dec 98\\
 962 & C & 99 & 13.3 & 1.97 &  0.00 & 3.623 &    0.80 &   1.8 &  9.56 &   E & $\leq$11.0 & .... & .. &        Dec 98\\
 969 & C & 99 & 14.4 & 2.54 & $-$0.54 & 3.544 &    0.58 &  $-$2.7 &  .... & ... &  14.0 $\pm$  1.8 & .... & .. &        Dec 98\\
 971 & C & 83 & 11.5 & 1.51 &  0.40 & 3.602 &    0.19 &   2.4 &  .... & ... &  42.7 $\pm$  3.6 & .... & .. &        Dec 98\\
 972 & P & 99 & 13.3 & 2.02 & $-$0.28 & 3.562 &    0.38 &   1.2 &  8.07 &   V &  11.1 $\pm$  2.1 & 1.77 & .. &        Dec 97\\
 988 & C & 99 & 14.8 & 3.12 & $-$0.34 & 3.544 &    0.81 &  $-$4.2 &  .... & ... &  19.9 $\pm$  6.1 & .... & .. &        Dec 97\\
 990 & C & 99 & 14.1 & 2.88 & $-$0.42 & 3.500 &   $-$0.14 &   0.6 &  2.16 &   E &  21.4 $\pm$  3.6 & 0.91 & .. &        Dec 98\\
 995 & C & 98 & 13.0 & 2.11 & $-$0.14 & 3.555 &    0.22 &   0.9 &  .... & ... & $\leq$11.0 & .... & .. &        Jan 97\\
 998 & C & 99 & 13.2 & 2.32 & $-$0.24 & 3.535 &    0.11 &  .... &  .... & ... &  12.8 $\pm$  1.6 & .... & .. &        Dec 98\\
1002 & C & 99 & 14.6 & 3.08 & $-$0.41 & 3.471 &    0.51 & $-$18.6 &  .... & ... & ........ & .... & .. &        Dec 98 \\
1004 & C & 99 & 13.8 & 2.10 & $-$0.40 & 3.562 &    0.31 &   2.0 &  .... & ... & $\leq$11.0 & .... & .. &        Dec 98\\
1006 & C & 99 & 14.7 & 2.92 & $-$0.65 & 3.518 &    2.21 &   1.2 &  .... & ... & $\leq$11.0 & .... & .. &        Dec 98\\
1007 & C & 99 & 14.5 & 2.50 & $-$0.72 & 3.526 &   $-$0.00 &   1.1 &  .... & ... &  11.1 $\pm$  1.5 & .... & .. &        Dec 98\\
1008 & C & 92 & 12.7 & 1.71 & $-$0.08 & 3.571 &    0.03 &   2.0 &  .... & ... & $\leq$11.0 & .... & .. &        Jan 97\\
1013 & C & 99 & 14.6 & 2.30 & $-$0.74 & 3.526 &    0.64 & $-$26.5 &  .... & ... &  25.6 $\pm$  5.7 & .... & .. &        Dec 98\\
1016 & C & 88 & 13.0 & 1.92 &  0.16 & 3.643 &    0.38 &  .... &  .... & ... &  12.4 $\pm$  1.0 & .... & .. &        Dec 98\\
1020 & C & 99 & 12.7 & 1.40 & $-$0.08 & 3.623 &   $-$0.01 &   1.7 &  .... & ... & $\leq$11.0 & .... & .. &        Jan 97\\
1021 & C & 36 & 13.0 & 1.91 &  0.11 & 3.633 &    0.34 &   1.6 &  7.75 &   E &  15.0 $\pm$  1.7 & 2.30 & .. &        Dec 97\\
1026 & C & 98 & 13.8 & 2.24 & $-$0.27 & 3.571 &    1.41 &  $-$1.3 &  4.21 &   E &  15.3 $\pm$  1.7 & 1.28 & .. & Jan 97,Dec 97\\
1030 & C & 99 & 13.7 & 2.18 & $-$0.43 & 3.544 &    0.36 &   1.5 &  .... & ... &  11.9 $\pm$  1.4 & .... & .. &        Dec 98\\
1033 & C & 80 & 14.7 & 2.04 &  0.49 & 3.917 &   $-$0.31 &   2.0 &  .... & ... &  34.2 $\pm$  5.7 & .... & .. &        Dec 98\\
1035 & C & 98 & 14.1 & 2.35 & $-$0.55 & 3.526 &    0.09 &   1.6 &  0.70 &   E &  77.2 $\pm$ 14.7 & 1.07 & .. &        Dec 98\\
1040 & C & 99 & 13.6 & 2.66 & $-$0.31 & 3.518 &    0.20 &   0.6 &  .... & ... &  11.7 $\pm$  2.1 & .... & .. &        Dec 97\\
1042 & C & 99 & 16.0 & 2.70 & $-$1.08 & 3.483 &    0.55 &   0.0 &  .... & ... & ........ & .... & .. &        Dec 97 \\
1044 & C & 87 & 14.6 & 3.30 & $-$0.54 & 3.500 &    0.23 &  $-$1.5 &  2.98 &   E & ........ & .... & .. &        Dec 98 \\
1047 & C & 99 & 14.2 & 2.84 & $-$0.51 & 3.509 &   $-$0.03 &   0.6 &  .... & ... &  16.3 $\pm$  2.8 & .... & .. &        Dec 97\\
1049 & C & 95 & 11.8 & 1.57 &  0.30 & 3.612 &    0.10 &   2.2 &  .... & ... &  18.6 $\pm$  1.2 & .... & .. &        Jan 97\\

\enddata
\protect\label{table:all data}
\end{deluxetable}

\clearpage
\begin{deluxetable}{ccccc}
\tablewidth{0pt}
\tablecaption{Gliese Stars Used as Narrow-lined Spectral Templates}
\tablehead{\colhead{Gliese\#} & \colhead{Spectral Type} & \colhead{$V$} & \colhead{RA (1950)} & \colhead{Dec (1950)}\\
\colhead{} & \colhead{} & \colhead{(mag)} & \colhead{(h m s)} &
\colhead{($\deg$ ' '')}}
\startdata
75 & K0V & 5.6 & 01 44 06 & +63 36 24 \\
144 & K2V & 3.7 & 03 30 34 & $-$09 37 36 \\
114 & K7V & 8.9 & 02 47 49 & +15 30 36 \\
15A & M1.5V & 8.1 & 00 15 31 & +43 44 24 \\
411 & M2V & 7.5 & 11 00 37 & +36 18 18 \\
\enddata
\protect\label{table:templates}
\end{deluxetable}

\clearpage
\begin{deluxetable}{lrcrccccc}
\tablewidth{0pt}
\tabletypesize{\small}
\tablecaption{Mean Radii}
\tablehead{\colhead{Bin} & \colhead{log~$L/L\sun$~range} &
\colhead{log~$T_{\rm{eff}}$~range} & \colhead{N} &
\colhead{$\langle$$Rsin(i)$$\rangle$} & \colhead{$\langle Rsin(i)
\rangle \over 0.79$} & \colhead{$R$ for bin center}\\
\colhead{} & \colhead{} & \colhead{} & \colhead{} &
\colhead{(R$_{\sun}$)} & \colhead{(R$_{\sun}$)} & \colhead{(R$_{\sun}$)}}
\startdata
A1 & 1.11 $-$  1.48 & 3.682 $-$ 3.730 & 3   &  3.53 $\pm$  0.72   &   4.49 $\pm$  0.92  &   5.75\\ 
A2 & 0.74 $-$  1.11 & 3.682 $-$ 3.730 & 5   &  2.85 $\pm$  0.32   &   3.62 $\pm$  0.41  &   3.75\\ 
\\
B1 & 1.11 $-$  1.48 & 3.634 $-$ 3.682 & 2   &  3.86 $\pm$  0.12   &   4.91 $\pm$  0.15  &   7.17\\ 
B2 & 0.74 $-$  1.11 & 3.634 $-$ 3.682 & 3   &  3.34 $\pm$  0.10   &   4.25 $\pm$  0.13  &   4.68\\ 
B4 & 0.00 $-$  0.37 & 3.634 $-$ 3.682 & 3   &  1.18 $\pm$  0.37   &   1.51 $\pm$  0.47  &   2.00\\ 
\\
C3 & 0.37 $-$  0.74 & 3.586 $-$ 3.634 & 2   &  2.64 $\pm$  0.53   &   3.36 $\pm$  0.67  &   3.81\\ 
C4 & 0.00 $-$  0.37 & 3.586 $-$ 3.634 & 3   &  1.16 $\pm$  0.36   &   1.48 $\pm$  0.45  &   2.49\\ 
\\
D4 & 0.00 $-$  0.37 & 3.538 $-$ 3.586 & 8   &  1.52 $\pm$  0.16   &   1.94 $\pm$  0.21  &   3.11\\ 
D5 & $-$0.37 $-$  0.00 & 3.538 $-$ 3.586 & 10  &  1.23 $\pm$  0.11   &   1.56 $\pm$  0.14  &   2.03\\ 
\\
E5 & $-$0.37 $-$  0.00 & 3.490 $-$ 3.538 & 10  &  1.08 $\pm$  0.11   &   1.37 $\pm$  0.14  &   2.53\\ 
E6 & $-$0.74 $-$ $-$0.37 & 3.490 $-$ 3.538 & 10  &  0.92 $\pm$  0.08   &   1.17 $\pm$  0.10  &   1.65\\ 
\enddata
\protect\label{table:radii}
\end{deluxetable}
\end{document}